\documentstyle[12pt,a4,epsf]{article}
\newcommand{\rcite}[1]{{\cite{#1}}}
\newcommand{\rref}[1]{{(\ref{#1})}}
\newcommand{\tref}[1]{{\ref{#1}}}
\newcommand{\rlabel}[1]{{\label{#1}}}
\newcommand{\rbibitem}[1]{\bibitem{#1}}
\newcommand{\be}{\begin{equation}}
\newcommand{\ee}{\end{equation}}
\newcommand{\ba}{\begin{eqnarray}}
\newcommand{\ea}{\end{eqnarray}}
\newcommand{\dis}{\displaystyle}

\newcommand{\mathrm}[1]{{\rm #1}}
\newcommand{\tr}{\mathrm{tr}}
\def\theequation{\arabic{section}.\arabic{equation}}

\begin{document}
\begin{titlepage}
\begin{flushright}
{FTUV/96-69}\\
{IFIC/96-78}\\
{NORDITA-96/70 N,P}\\
{hep-ph/9610360}\\
Revised
\end{flushright}
\vspace{2cm}
\begin{center}
{\large\bf Electromagnetic Corrections for Pions and Kaons :
Masses and Polarizabilities}\\
\vfill
{\bf Johan Bijnens$^a$ and Joaquim Prades$^{b}$}\\[0.5cm]
${}^a$ NORDITA, Blegdamsvej 17,\\
DK-2100 Copenhagen \O, Denmark\\[0.5cm]
$^b$ Departament de F\'{\i}sica Te\`orica, Universitat de Val\`encia
\\and IFIC, CSIC - Universitat de Val\`encia, C/ del Dr. Moliner 50,
E-46100 Burjassot (Val\`encia) Spain
\end{center}
\vfill
\begin{abstract}
The unknown constants in Chiral Perturbation Theory needed for
an all orders analysis of the polarizabilities and
electromagnetic corrections to the masses of
the pseudo-Goldstone bosons are estimated at
leading order in $1/N_c$. We organize the calculation in an
$1/N_c$-expansion and separate long- and short-distance physics
contributions by introducing an Euclidean cut-off.
The long-distance part is evaluated using the ENJL model
and the short-distance part using perturbative QCD and factorization.
We obtain very good matching between both.

We then include these estimates in a full Chiral Perturbation  Theory
calculation to order $e^2$ $p^2$ for the masses
and $p^6$ for the polarizabilities. For the electromagnetic
corrections to the masses,  we confirm a
large violation of Dashen's theorem getting a more
precise value for this violation. We make comparison with
earlier related work. Some phenomenological
consequences are discussed too.
\end{abstract}
\vspace*{1cm}
\vfill
\noindent October 1996
\end{titlepage}

\section{Introduction}
\rlabel{first}
\setcounter{equation}{0}

Virtual electromagnetic (EM) effects in purely strong processes can be
important in precision situations. This is especially true in
the case of isospin breaking contributions to hadron masses
and some hadronic processes.
If we want a high precision description of the latter, we need
to know not only the effects due to the quark mass difference $m_d-m_u$,
which is a quantity we would also like to extract from these experiments,
but the size of the electromagnetic contributions.

That these contributions can be sizeable in certain cases
is best illustrated in the case of the
observed $\pi^+-\pi^0$ mass difference which is almost entirely due
to photon loops \rcite{das}.
At present, we cannot directly use QCD to estimate these effects.
Some first progress using lattice QCD has been made recently
in \rcite{DET}.
Instead we turn to the method of Chiral Perturbation Theory (CHPT). In
the purely mesonic sector for the strong and semi-leptonic processes,
this was started as a systematic program by Gasser and
Leutwyler \rcite{GL84,GL85} and has since then
been extended to a large variety of
processes\rcite{reviews}. In the case of
$\gamma\gamma\to\pi\pi$\rcite{BGS94,buergi}, $\pi\pi$-scattering
\rcite{pipi} and $\pi \to \ell \nu \gamma$ \rcite{pere} this
has even been done to the two-loop order.
In $\pi\pi$-scattering, electromagnetic
corrections might also become relevant
at the precision achieved by the two-loop order calculation.
They are already important, at the present level of precision,
in various others form factors.

Chiral Perturbation Theory for virtual
electromagnetic effects was first described
at lowest order ($e^2 p^0$) in \rcite{dashen}.
Urech\rcite{urech} has recently systematically studied the
next-to-leading order terms. His work has been mainly
dedicated to the EM correction in the masses of the lowest
pseudoscalar Goldstone bosons. This program  has been
later expanded into a few more form factors by
Neufeld and Rupertsberger\rcite{neru,neru96}.
The results were, however, quite
dependent on the values for the new coupling constants appearing at
order $e^2$ $p^2$ in the chiral expansion.
Unfortunately, contrary to the case of strong and semileptonic processes,
it is impossible to determine all these constants from experimental
data. The calculation of these constants is the main subject of this paper.
In addition we provide estimates for the counterterms appearing
in CHPT at order $p^6$ in the processes $\gamma \gamma \to PP$.
Here we extend the work done in \rcite{BFP96} to
the charged case and make predictions for the polarizabilities
of pions and kaons too. This side aspect is discussed
in Section \tref{polarizabilities}.

The real problem in calculating
the constants is that it requires an integration
over internal photon momenta. For instance, in the estimate
done for the strong sector in \rcite{EGPR89} using
the lowest lying
resonance saturation,  only their lowest order couplings are relevant.
When one integrates over all photon
momenta, the couplings of the resonances to all orders need to be known,
 thus making these estimates much more difficult.
The lowest order constant, called $C$ in Section \tref{second},
 has a very long history, it was first estimated
 in \rcite{das} using PCAC and then saturating  two-point functions
with resonance exchanges.
A different technique was subsequently developed by
Bardeen, Buras and G\'erard in
\rcite{BBG87} for weak non-leptonic matrix elements, now
generally referred to as the $1/N_c$-approach. This approach was
then used together with the Das et al. sum rule \rcite{das}
to estimate the
pion mass difference at lowest order, or equivalently $C$,
 in \rcite{BBG89}.  Here a proper
separation of long and short distance contributions was also possible.
The same method has been used in the chiral quark
model\rcite{BR91} and in the extended Nambu--Jona-Lasinio (ENJL)
 model\rcite{BRZ94}. All of these only
allowed estimates of the lowest order constant $C$ since
they were based on the Das et al. sum rule.

The calculation of the pion and kaon EM mass differences
beyond lowest order in CHPT, using saturation by
the lowest lying resonances, was recently performed
 by Donoghue, Holstein, and Wyler in \rcite{DHW}
 and Baur and Urech in \rcite{bau}. For earlier attempts see
\rcite{socolow}.
In \rcite{logs} the  chiral logs at $\nu=1$ GeV
were used to estimate these EM mass differences.
These papers all had to make assumptions about the short distance behaviour.
More comments about these assumptions are in Section \tref{comparison}.
The short-distance contribution was
introduced in an operator-product expansion (OPE)
framework in \rcite{hans93}. This is
discussed in Section \tref{sd}.

The method used in this paper is an extension of the original $1/N_c$
method\rcite{BBG87}. We use an off-shell Green function. This
method was used by us previously in the calculation of
the $\Delta S =2$ hadronic matrix element in the $K^0$-$\overline
{K^0}$ system and commonly parametrized by the so-called
$B_K$ factor \rcite{BP95}.

The paper is organized as follows: in Section \tref{second} we
shortly discuss CHPT for electromagnetic corrections and define our
set of counterterms. Our set is somewhat
 more appropriate to the large $N_c$
limit and different number of flavours. We also point out in some
detail the gauge dependence of the generating functional. No observable
quantities do of course depend on the gauge but the infinite parts of
the constants at next-to-leading order do depend on it. In Section
\tref{method} we explain the method. This section also gives
a short overview of the ENJL model that we use for the
long-distance contributions. The main contributions are
the latter due to the photon propagator.
In Section \tref{results}
we give the results and use CHPT at leading order in $1/N_c$ to
extract the CHPT constants. We compare with the earlier
work in Section \tref{comparison}.
 Section \tref{quarkmasses} summarizes the consequences for the
ratios of the light current quark masses.
 Section \tref{polarizabilities} contains the
results on pion and kaon polarizabilities and we present our main
conclusions in Section \tref{conclusions}.

\section{Chiral Perturbation Theory  Analysis}
\rlabel{second}
\setcounter{equation}{0}

In this Section we use CHPT to analyse the two-point functions,
\be
\rlabel{twopoi}
 \Pi(q^2) = i {\dis \int} \, {\rm d}^4 x \, e^{iqx} \,
 \langle 0 | T \left( P(0) P^\dagger(x) \right) | 0 \rangle\,,
\ee
in the presence of electromagnetic interactions.
We shall do this  to first order in $\alpha_{QED}$ and to order $p^2$
in CHPT. The pseudoscalar source $P(x)$ in \rref{twopoi} is
defined as $\left( \overline q_a i\gamma_5 q_b\right)(x)$,
with $a$, $b$ quark flavour indices and colour indices summed inside
the parenthesis.

 At lowest order in the chiral expansion ${\cal O} (p^2)$ \rcite{wein},
the strong interactions between the lowest pseudoscalar mesons including
external vector, axial-vector, scalar and pseudoscalar sources
are described by the following effective Lagrangian
\ba
\rlabel{lowest}
{\cal L}^{(2)}_{\rm eff}& =& \frac{\dis F_0^2}{\dis 4}
\left\{ \tr \left( D_\mu U D^\mu U^\dagger \right) + \tr
\left( \chi U^\dagger + U \chi^\dagger \right) \right\}
\ea
where $D_\mu$ denotes the covariant derivative

\be
\rlabel{cova}
D_\mu U = \partial_\mu U -i (v_\mu + a_\mu)U
+ i U (v_\mu-a_\mu),
\ee
and $U \equiv {\rm exp} \left( \frac{\dis i \sqrt 2 \Phi}{\dis F_0}
 \right)$
an SU(3) matrix incorporating the octet of pseudoscalar mesons
\be \rlabel{Uoctet}
\Phi(x)=\frac{\vec{\lambda}}{\sqrt 2} \vec{\phi} =
\left( \begin{array}{ccc}
\frac{\dis \pi^0}{\dis \sqrt 2} + \frac{\dis \eta_8}
{\dis \sqrt 6} &  \pi^+ & K^+ \\
\pi^- & -\frac{\dis \pi^0}{\dis \sqrt 2}+\frac{\dis \eta_8}
{\dis \sqrt 6} & K^0 \\
K^- & \overline K^0 & -\frac{\dis 2 \eta_8}{\dis \sqrt 6} \end{array}
\right) \, .
\ee
In  Eq. \rref{cova}, $v_\mu(x)$ and $a_\mu(x)$ are external 3 $\times$ 3
vector and axial-vector field matrices. When electromagnetism is
switched on, $v_\mu(x)=|e| Q A_\mu(x)$ and $a_\mu(x)=0$.
Here $A_\mu(x)$ is the photon field and the light-quark electric
charges in units of the electron charge $|e|$ are collected in
the 3 $\times$ 3 flavour matrix $Q = \frac{1}{3} {\rm diag}(2,-1,-1)$.
 In Eq. \rref{lowest} $\chi \equiv 2 B_0 ({\cal M} +s(x)+ i p(x))$
with $s(x)$ and $p(x)$  external scalar and pseudoscalar 3
$\times$ 3 field matrices and ${\cal M}$ the 3 $\times$ 3
flavour matrix ${\cal M} = {\rm diag}(m_u,m_d,m_s)$
collecting 
the light-quark current masses. The constant $B_0$ is related to
the vacuum expectation value
\be
\left. \langle 0| \bar q q | 0 \rangle \right|_{q=u,d,s} =
- F_0^2 B_0 \left(1 + {\cal O}( m_q)\right) \, .
\ee

In this normalization, $F_0$ is the chiral limit value corresponding
to the pion decay coupling $F_\pi \simeq 92.4$ MeV.
In the absence
of the U(1)$_A$ anomaly (large $N_c$ limit) \rcite{anoma}, the
SU(3) singlet $\eta_1$ field becomes the ninth Goldstone boson
which is incorporated in the $\Phi(x)$ field as
\be
\rlabel{eta1}
\Phi(x)= \frac{\vec{\lambda}}{\sqrt 2} \vec{\phi} + \frac{\eta_1}{\sqrt 3}
{\mbox {\large \bf 1}} \, .
\ee

To this order, the two-point functions in \rref{twopoi}
have the following form
\be
\rlabel{chpttwo}
\Pi(q^2)= -\frac{2 B_0^2 F_0^2}{q^2-m_0^2},
\ee
where $m_0$ is the pseudo-Goldstone boson masses to that order, i.e.
$m_0^2=B_0 (m_a+m_b)$, with $m_a$ the flavour $a$ quark mass.
Here we are interested in the isospin breaking corrections
to the poles of these two-point functions
induced by electromagnetism to order $e^2$.
So we will set $m_u=m_d$ in the calculation.

To take into account virtual photons excitations we need
to add to the Lagrangian in \rref{lowest} the corresponding
kinetic and gauge fixing terms, i.e.
\be
\rlabel{kinetic}
  - \frac{1}{4} F^{\mu\nu}F_{\mu \nu} -
\frac{1}{2(1-\xi)} \, \left( \partial_\mu A^\mu \right)^2
\ee
where $F^{\mu\nu}= \partial^\mu A^\nu - \partial^\nu A^\mu$
is the electromagnetic field strength tensor. The parameter $\xi$ is
the gauge fixing parameter which is zero in the Feynman
gauge, one in the Landau gauge and four in Yennie's one.

\subsection{Lowest Order Contribution}
\rlabel{lowestorder}

At lowest order in the chiral expansion, electromagnetic (EM)
virtual interactions  of order $e^2$ between the pseudo-Goldstone
bosons are described by the following effective Lagrangian
\rcite{GL85,anoma,PR95}
\ba
\rlabel{e2lowest}
{\cal L}^{(0)}_{e^2} &=& e^2 C_1(\Phi_0^2) \tr \left(Q^2 \right) +
e^2 C_2(\Phi_0^2) \tr \left(Q U Q U^\dagger \right) .
\ea
Here the $U$ field matrix is the U(3) symmetric one in \rref{eta1}
and $\Phi_0= \theta + \eta_1 \sqrt 6 / F_0$,
where $\theta$ is the so-called QCD theta-vacuum parameter.
The constant $C=C_2(0)$ is the coupling
introduced in \rcite{EGPR89}.
There are no loop contributions to this order and therefore
$C_{1,2}(0)$ and the derivatives of $C_{1,2}(x)$ at $x=0$ are
finite counterterms not fixed by symmetry alone.
In the large $N_c$ limit, $C_{1,2}(0)$ are of order $N_c$ whereas
the $n$-th derivative of $C_{1,2}(x)$ is of order $1/N_c^{2n-1}$.
To  order $e^2 p^0$, the correction to the pole position
of the two-point function in \rref{chpttwo} is zero
for $P=\pi^0$ and $P=\eta_8$, while charged pion and kaons
get the same non-zero correction, namely
\be
m^2_{\em EM (\pi^+,K^+)}= e^2 \, \frac{2 C}{F_0^2}.
\ee
This is the so-called Dashen's theorem \rcite{dashen}.
The pole position for the $\eta_1$-$\eta_1$ two-point function
gets corrected by
\be
m^2_{\em EM (\eta_1)}= e^2 \, \frac{8 (C'_1(0)+C'_2(0))}{F_0^2}
\ee
from EM virtual interactions.

\subsection{Next-to-Leading Order Contribution}
\rlabel{next}

In this Section we shall report on the CHPT order $p^2$ $e^2$
virtual EM corrections. The first type of these corrections
is the emission and absorption of a photon by a pseudo-Goldstone
boson line (see Fig. \tref{fig1}).  To the order we are interested here,
the $\gamma P^+ P^-$ and $\gamma \gamma P^+ P^-$ vertices come
 from the Lagrangian in \rref{lowest}.
 This contribution needs a counterterm
of order $e^2 p^2$ to make it UV finite.
\begin{figure}
\begin{center}
\leavevmode\epsfxsize=12cm\epsfbox{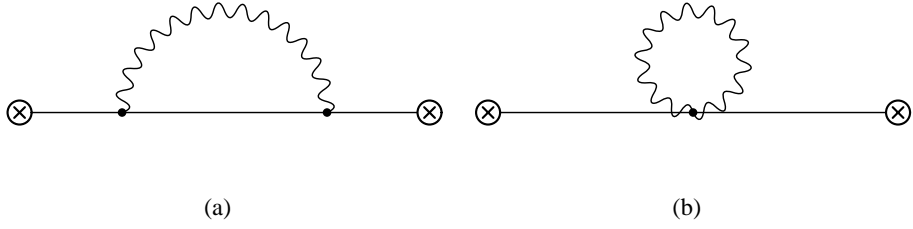}
\end{center}
\caption{\rlabel{fig1}The photon loop contributions to
\protect{\rref{empoint}}. The crosses are pseudoscalar currents. The wiggly
line is the photon. The full lines are pseudoscalars.}
\end{figure}

The complete CHPT order $p^2$ effective Lagrangian describing virtual EM
interactions of order $e^2$ between pseudo-Goldstone bosons and external
vector, axial-vector, scalar and pseudoscalar sources was written
in \rcite{urech} (see also \rcite{neru96}).  The coefficients of this
Lagrangian are the needed counterterms to absorb the UV
divergences of order $e^2 p^2$. To construct this
Lagrangian, some Cayley-Hamilton relations for SU(3) matrices were used.
We would like here
to include also the ninth pseudo-Goldstone boson as above
and work with U(3) matrices (see Eq. \rref{eta1}),
which is the symmetry in the
large $N_c$ limit ($N_c$ is the number of colours).
 This is useful for our calculation since we want to use
the $1/N_c$-expansion as the organizing
scheme. To order $e^2 p^2$, one has to add
to the Lagrangians in \rref{lowest}, \rref{kinetic}, 
and \rref{e2lowest} the following one:
\ba
\rlabel{e2next}
{\cal L}^{(2)}_{e^2} & = &
e^2 \tilde K_1(\Phi_0^2) F_0^2 \, \tr \left( D^\mu U^\dagger D_\mu U \right)
\tr \left( Q^2 \right) \nonumber \\  &+&
e^2 \tilde K_2(\Phi_0^2) F_0^2 \, \tr \left( D^\mu U^\dagger D_\mu U \right)
\tr \left( Q U Q U^\dagger \right) \nonumber \\
& + & e^2 \tilde K_3 (\Phi_0^2)
F_0^2 \, \tr \left( Q U^\dagger  D^\mu U Q D_\mu
U^\dagger U + Q U D^\mu U^\dagger Q D_\mu U U^\dagger \right)
\nonumber \\ &+&  e^2 \tilde K_4(\Phi_0^2)
 F_0^2 \, \tr \left( Q U^\dagger D^\mu U Q D_\mu U U^\dagger
\right) \nonumber \\
& + & e^2 \tilde K_5(\Phi_0^2)
 F_0^2 \, \tr \left( \left( D^\mu U^\dagger D_\mu U +
D^\mu U D_\mu U^\dagger \right) Q^2 \right) \nonumber \\ &+&
e^2 \tilde K_6(\Phi_0^2)
 F_0^2 \, \tr \left(D^\mu U^\dagger D_\mu U Q U^\dagger Q U +
D^\mu U D_\mu U^\dagger Q U Q U^\dagger \right) \nonumber \\
& + & e^2 K_7(\Phi_0^2)
 F_0^2 \, \tr \left( \chi^\dagger U + U^\dagger \chi \right)
\tr \left( Q^2 \right) \nonumber \\
&+&  e^2 K_8(\Phi_0^2)  F_0^2 \, \tr
\left( \chi^\dagger U + U^\dagger \chi \right) \tr \left( Q U Q U^\dagger
\right) \nonumber \\ &+& e^2 K_9(\Phi_0^2) F_0^2 \, \tr \left(
\left( \chi^\dagger U + U^\dagger \chi + \chi U^\dagger +
U \chi^\dagger \right) Q^2 \right) \nonumber \\ & + &
e^2 K_{10}(\Phi_0^2)
 F_0^2 \, \tr \left( \left( \chi^\dagger U + U^\dagger \chi \right)
Q U^\dagger Q U + \left( \chi U^\dagger + U \chi^\dagger \right)
 Q U Q U^\dagger \right) \nonumber \\
& + & e^2
K_{11}(\Phi_0^2)
 F_0^2 \, \tr \left( \left( \chi^\dagger U - U^\dagger \chi \right)
Q U^\dagger Q U + \left( \chi U^\dagger - U \chi^\dagger \right)
Q U Q U^\dagger \right)  \nonumber \\
& + & e^2 K_{12}(\Phi_0^2) F_0^2 \, \tr \left( U D^\mu U^\dagger
 \left[ \nabla^R_\mu Q , Q \right] \, + \,  U^\dagger D^\mu U
\left[ \nabla^L_\mu Q , Q \right] \right) \nonumber \\
& + & e^2 K_{13}(\Phi_0^2)
 F_0^2 \, \tr \left( \nabla^R_\mu Q U \nabla_L^\mu  Q U^\dagger \right)
\nonumber \\ &+&  e^2 K_{14}(\Phi_0^2)
 F_0^2 \, \tr \left( \nabla^R_\mu Q \nabla_R^\mu Q +
\nabla^L_\mu Q \nabla_L^\mu Q \right) \nonumber \\
&+& e^2 \tilde K_{15}(\Phi_0^2) F_0^2 \tr \left( \left( D_\mu U
 U^\dagger - D_\mu U^\dagger U \right) Q^2 \right) \,
\delta^\mu \theta \nonumber \\
&+& e^2 \tilde K_{16}(\Phi_0^2) F_0^2 \tr \left( D_\mu U Q
 U^\dagger Q - D_\mu^\dagger Q U Q \right)  \, \delta^\mu \theta
\nonumber \\
&+& e^2 \tilde K_{17}(\Phi_0^2) F_0^2
 \left( \tr \left( Q^2 \right) \, + \,
 \tr \left( Q U Q U^\dagger \right) \right) \,
D^\mu \Phi_0 D_\mu \Phi_0  \nonumber \\
&+& e^2 \tilde K_{18}(\Phi_0^2) F_0^2
 \left( \tr \left( Q^2 \right) \, + \,
 \tr \left( Q U Q U^\dagger \right) \right) \,
D^\mu \Phi_0 \delta_\mu \theta  \nonumber \\
&+& e^2 \tilde K_{19}(\Phi_0^2) F_0^2
 \left( \tr \left( Q^2 \right) \, + \,
 \tr \left( Q U Q U^\dagger \right) \right) \,
\delta^\mu \theta \delta_\mu \theta  \nonumber \\
&+&  e^2 \Phi_0 \tilde K_{20} (\Phi_0^2) F_0^2  \tr \left(
\chi^\dagger U - U^\dagger \chi \right)
\tr \left(Q^2 \right) \nonumber \\
&+&  e^2 \Phi_0 \tilde K_{21} (\Phi_0^2) F_0^2 \tr \left(
\chi^\dagger U - U^\dagger \chi \right) \tr \left( Q U Q U^\dagger
\right) \nonumber \\
&+& e^2 \Phi_0 \tilde K_{22}(\Phi_0^2) F_0^2 \tr \left(
\left( U \chi^\dagger + \chi^\dagger U - U^\dagger \chi -
\chi U^\dagger \right) Q^2 \right) \nonumber \\
& + & e^2 \Phi_0 \tilde K_{23}(\Phi_0^2)
 F_0^2 \tr \left( \left( \chi^\dagger U
 + U^\dagger \chi \right) Q U^\dagger Q U - \left( \chi U^\dagger +
U^\dagger \chi \right) Q U Q U^\dagger \right) \nonumber \\
& + & e^2 \Phi_0 \tilde K_{24}(\Phi_0^2)
 F_0^2 \tr \left( \left( \chi^\dagger
U - U^\dagger \chi \right) Q U^\dagger Q U - \left( \chi U^\dagger
- U \chi^\dagger \right) Q U Q U^\dagger \right) \nonumber \\
\ea
It is interesting to study the large $N_c$
behaviour of the coefficients in \rref{e2next}. In the large $N_c$ limit,
the couplings $\tilde K_1(0)$, $\tilde
K_2(0)$, $K_7(0)$, $K_8(0)$, $\tilde K_{15}(0)$, $\tilde K_{16}(0)$,
$\tilde K_{22}(0)$, $\tilde K_{23}(0)$, and $\tilde K_{24}(0)$
are order $1/N_c$ and the couplings
$\tilde K_{17}(0)$, $\tilde K_{18}(0)$, $\tilde K_{19}(0)$,
$\tilde K_{20}(0)$, and  $\tilde K_{21}(0)$ are order $1/N_c^2$.
The rest of the constants $K_i(0)$ and
$\tilde K_i(0)$ in \rref{e2next} are of order 1. Each derivative of 
the $K_i(x)$ and  $\tilde K_i(x)$ 
functions brings in an additional factor $1/N_c^2$.
In the rest of the paper, we call the  functions
$K_i(\Phi_0^2)$ and $\tilde K_j(\Phi_0^2)$
at $\Phi_0=0$ the couplings $K_i$ and $\tilde K_j$.

 The covariant derivatives $\nabla^{L,R}_\mu$ are defined as
follows,
\be
\nabla^{L(R)}_\mu Q = \partial_\mu Q - i [v_\mu-(+)a_\mu, Q] ,
\ee
and the symbol $\delta_\mu \theta$ is
\be
\delta_\mu \theta \equiv \partial_\mu \theta + 2 \tr \left(a_\mu
\right) .
\ee

 The relation between these couplings and the ones defined in
\rcite{urech,neru96} is
\ba
\rlabel{trans}
K_1 = \tilde K_1 + \tilde K_3 &;& K_2 = \tilde K_2 +
    \frac{\tilde K_4}{2}
\nonumber \\ K_3 = - \tilde K_3 &;& K_4 =  \tilde K_4 \nonumber \\
K_5 = \tilde K_5 - 2 \tilde K_3 &;& K_6 = \tilde K_6 -  \tilde K_4 .
\ea
{}From here one gets the
following large $N_c$ limit relations
\be
K_1 = - K_3 \hspace*{1cm} {\rm and}  \hspace*{1cm} 2 K_2 = K_4 .
\ee
The rest of the tilded couplings introduced here were not included
in those references because they worked in an octet symmetry framework.
The expressions of the corrections to the poles
of pseudoscalar two-point functions at this order are given in
Appendix \tref{A}. The coefficients of the effective Lagrangian in
\rref{e2next} are again not
determined from chiral symmetry arguments alone. Its estimation
is the central subject of this work. We explain the technique we use
in the next section.

\subsection{Gauge Dependence of the Various Quantities}
\rlabel{gauge}

A subtle issue is involved here. The generating functional in terms of
colourless external fields, as used in \rcite{GL85}, is not independent
of the gauge chosen for the gauge fields propagators
(in particular the photon one).
As a well known consequence
Green  functions are not gauge invariant in general.
The underlying reason is simply that external sources
are in general charged so they transform under the gauge group
non-trivially. Of course, for observable
quantities, like the mass shift we obtain from the two-point function
\rref{twopoi} or any other physical quantity, the result has to be
gauge invariant. So this gauge dependence disappears when the external
sources are on the mass-shell.

For instance, one obtains a gauge dependence in
the result for the two-point function \rref{twopoi}
which  only cancels when the meson created by $P^\dagger(x)$
and destroyed by $P(x)$ is on the mass-shell.
This means that some of the couplings $K_i$ and $\tilde K_i$
are actually U(1) gauge dependent. In practice,
 in our CHPT calculation 
we  fix the gauge for the photon propagator
to be the Feynman one  (i.e. $\xi=0$).
The same gauge was used in the CHPT calculations
 in \rcite{urech,neru96}. If one wants to compare or use
the values of the couplings we get with the ones
obtained from experiment or other model calculations, one
should make sure that the Feynman gauge (the gauge we used)
is used in the CHPT calculation or in the model calculation.
Alternatively, one can of course compare directly the same physical
quantity.

This gauge dependence
does not reduce the number of parameters in the CHPT Lagrangian,
since choosing  a clever gauge fixing in order to remove a constant
from the Lagrangian would bring back the
parameter in the photon propagator.

\section{Calculation of the Counterterms}
\rlabel{method}
\setcounter{equation}{0}

We calculate directly
the two-point function defined in \rref{twopoi} in the presence of
EM interactions to order $e^2$. In practice, this means the
calculation of
\ba
\rlabel{empoint}
\Pi(q^2) = i \frac{e^2}{2} {\dis \int} \, \frac{{\rm d}^4 r}
{(2 \pi)^4} \, \frac{g_{\mu\nu} - \xi r_\mu r_\nu/r^2}
{r^2 -i \epsilon} \, \Pi^{\mu\nu}_{PPVV}(q,r).
\ea
Where $\Pi^{\mu\nu}_{PPVV}(q,r)$ is the following four-point
function
\be
\rlabel{fourpoint}
\Pi^{\mu\nu}_{PPVV}(q,r) \equiv
i^3  \int {\rm d}^4 x \, \int {\rm d}^4 y \, \int {\rm d}^4 z
\,  e^{-i(q x + r (y-z))} \,
\langle 0 | T  \left( P(0) P^\dagger(x) V^\mu(y) V^\nu(z)
\right) | 0 \rangle\, .
\ee
Here $V_\mu(x) = \left( \overline q (x) Q \gamma_\mu q(x) \right)$
with $\overline q(x)$
the SU(3) flavour vector $(\overline u(x), \overline
 d(x), \overline s(x))$.
Notice that since the photon momentum $r$ is integrated out, we should
know this function at all energies. In particular, this means
to all orders in a low energy CHPT expansion of this function
in the momentum $r$.
In order to extract as much information as possible
of the two-point function \rref{empoint},
 we calculate it at off-shell values of  $q^2$ as well.

Let us now  discuss on the U(1) gauge invariance of \rref{fourpoint}.
When the external pseudo-Goldstone bosons
are on-shell, the four-point function in \rref{fourpoint}
is U(1) gauge covariant and fulfills
\be
r_\mu \Pi^{\mu\nu}_{PPVV}(q,r) =
   r_\nu \Pi^{\mu\nu}_{PPVV}(q,r) = 0 .
\ee
Therefore the gauge dependent term proportional
to $\xi$ in \rref{empoint} cancels. This is no longer true when we move
to off-shell $q^2$ values. In that case, the  $\xi$ term gives a
non-zero contribution since we have no gauge covariance
for \rref{fourpoint} (see comments in Section \tref{gauge}).

The technique we use here is similar to the one introduced in
\rcite{BBG87} and used in \rcite{BBG89}, and is the variant
presented  in  \rcite{BP95}. This consists in introducing
an Euclidean cut-off in the integrated out photon momentum. This
cut-off $\mu$ both serves to separate long and short
distance contributions and as a matching variable.
After reducing the two pseudoscalar legs of
$\Pi(q^2)$ in \rref{twopoi} the result has no anomalous dimensions in
QCD. We have then to find a plateau in the cut-off $\mu$ if there
is good matching. The photon propagator will help to produce it.

So, after passing to Euclidean space Eq. \rref{empoint},
we introduce the cut-off $\mu$ in the photon momentum $r_E$ as follows
\ba
\rlabel{cut}
 {\dis \int_0^\infty} \, {\rm d} r_E
= {\dis \int_0^\mu} \, {\rm d} r_E \, +\,
{\dis \int_\mu^\infty} \, {\rm d} r_E .
\ea
For the short-distance part we can perform the full calculation,
see next Section.
In particular, we have obtained, in the large $N_c$ limit and
up to order $1/\mu^2$, the short-distance contributions
to all the terms in \rref{e2next} and not only those accessible via
\rref{empoint}.

\subsection{Short-Distance Contribution}
\rlabel{sd}

The higher part of the integral in \rref{cut} collects
the contributions of the higher than $\mu$ modes of
the virtual photons. The effective action obtained
by integrating out the virtual photons with modes higher
than $\mu$ in \rref{empoint}
can be expanded in powers of $1/\mu$ like in an OPE.
We compute these contributions up to order $1/\mu^2$.
This contribution was first  introduced in \rcite{hans93}.

There are four types of contributions.
There is a pure QED fermion mass renormalization contribution
(see Figure \tref{fig2}(a)) due to the fact that the quark mass in QED
runs proportionally
to the mass itself. The QED log divergency produces a contribution to the
effective action proportional to $\ln(\Lambda/\mu)$ where $\Lambda$
is the scale where the input current quark masses are renormalized in
QED. Within the ENJL model the input current quark masses are encoded
in the values of the constituent quark masses. We have fixed the
physical values for the constituent quark masses by comparing
the ENJL predictions to low energy observables typically
at some scale between the rho meson mass and the ENJL model cut-off
$\Lambda_{ENJL}=$ 1.16 {\rm GeV}. Accordingly,
 we will vary the scale $\Lambda$ in that range. Of course, there will
be some kind of double counting since we cannot disentangle
the EM virtual contributions to the ENJL parameters from the
experimental values. But they give order $e^4$ corrections.
\begin{figure}
\begin{center}
\leavevmode\epsfxsize=12cm\epsfbox{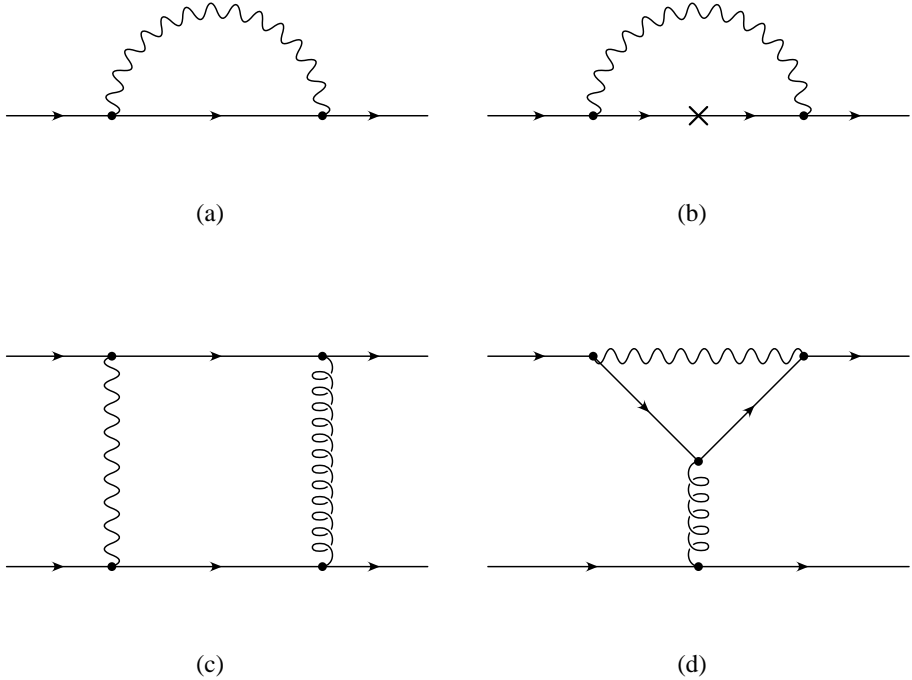}
\end{center}
\caption{\rlabel{fig2}The short distance contributions. The wiggly line
is the photon, the curly one the gluon, the full line a quark.
There are also the crossed and symmetric configurations.}
\end{figure}
The contribution of the QED self-energy diagram in Figure
 \tref{fig2}(a) to the effective QCD Lagrangian,
 in the presence of EM virtual interactions, is
\ba
\rlabel{SD1}
- \frac{3}{2} \,\frac{\alpha}{\pi}\,  \ln (\Lambda/\mu) \,
\left[ \left(\overline q_L \left( Q {\cal M} Q \right)^\dagger
 q_R \right)
+ \left( \overline q_R \left(Q {\cal M} Q\right)
  q_L \right) \right]\,
\ea
where colour indices are summed inside parenthesis.
Its effective realization can be calculated
to all orders in an $1/N_c$ expansion since it can be
written in terms of just bilinear QCD currents.
 At low energies, in terms
of the lowest pseudo-Goldstone bosons and external sources
and at lowest order in the chiral expansion, it only contributes
to the couplings $K_{10}$ and $K_{11}$ in \rref{e2next}.
\be
\rlabel{SD1had}
K_{10}^{\rm QED} = - K_{11}^{\rm QED} = \frac{3}{64 \pi^2} \,
\ln \left( \frac{\Lambda}{\mu} \right) \, .
\ee
Notice that here enters the scale $\Lambda$ where the current
quark masses are defined.
 The remaining part of the integration from $\Lambda$ to $\infty$
is absorbed in the definition of the {\em current} quark masses.
Here we can also indicate the type of corrections existing to
 \rref{SD1had}.
First, at the quark-gluon level there are $1/\mu^2$ and
$\alpha\alpha_S$ corrections to \rref{SD1}.
Then when going from the quark-gluon level expression 
in \rref{SD1} to the hadronic level  one, 
there are corrections that are higher order
in chiral power counting. To obtain \rref{SD1had} we used
the order $p^2$ strong Lagrangian
in \rref{lowest}. There are thus
CHPT ${\cal O}(p^4)$ corrections to this result.

The presence of explicit dependence on $\Lambda$ in the
short-distance contribution to $K_{10}$ and $K_{11}$ indicates
 that one has to be careful when using rules of the strong sector to
obtain naive estimates of order of magnitude
sizes of CHPT parameters. This we
will refer to later as failure of naive power counting.

This problem will appear whenever  non-leptonic couplings of
interactions other than the strong interaction come into play.
In particular it also shows up in
weak non-leptonic decays.
 There the effects are suppressed by extra inverse powers
of the $W$-boson mass, so its numerical importance is negligible.

A similar contribution  comes from diagram (b) in Fig. \tref{fig2}.
Here it is not the scalar and pseudoscalar current that are
renormalized but
the vector and axial-vector ones. They have to be defined at the scale
$\Lambda$ and again the QED running  can lead to a contribution.
The cross in Fig. \tref{fig2}(b) denotes an insertion of an 
external vector or axial-vector current. This will contribute to
$K_{12}$ and, with one more insertion, to
$K_{13}$ and $K_{14}$. There are in principle
short-distance contributions of
this type but they vanish because of the global chiral
invariance in QCD perturbation theory.
There will be short-distance contributions from this diagram
but to higher order operators in CHPT,
 for instance to  magnetic-like structures, etc.
So we have
\be
K^{\rm SD}_{12} = K^{\rm SD}_{13} = K^{\rm SD}_{14} = 0\,.
\ee
There can be long-distance contributions of this type due to
the spontaneous breaking of the axial symmetry.

The third and fourth type of contributions to the
effective Lagrangian are of order $N_c^2 \alpha_S$.
These are the well known box- and penguin-type diagrams
(see Figs. \tref{fig2}(c) and \tref{fig2}(d) and crossed versions).
To order $1/\mu^2$
they contribute to the effective Lagrangian as follows.
The box-type contribution is
\ba
\rlabel{box}
&& \frac{3}{2} \frac{\alpha \, \alpha^{\rm eff}_S(\mu^2)}{\mu^2} \,
 {\dis \sum_{a,b,c,d}}
\left[ \left(\overline q^a_L \gamma^\mu Q^{ab} q^d_L \right)
\left( \overline q^c_L \gamma_\mu Q^{cd} q^b_L \right) \, + \,
\left(\overline q^a_R \gamma^\mu Q^{ab} q^d_R \right)
\left( \overline q^c_R \gamma_\mu Q^{cd} q^b_R \right) \right.
    \nonumber \\
 &+& \left. 4 \, \left( \overline q^a_L Q^{ab} q^d_R \right)
\left( \overline q^c_R Q^{cd} q^b_L \right) \right] \, .
\ea
and the penguin-like contribution is
\ba
\rlabel{penguin}
&& \frac{1}{3} \frac{\alpha \, \alpha^{\rm eff}_S(\mu^2)}{\mu^2} \,
 {\dis \sum_{a,b,c}}
\left[ \left(\overline q^a_L \gamma^\mu q^c_L \right)
\left( \overline q^b_L \gamma_\mu (Q^2)^{bc} q^a_L \right) \, + \,
\left(\overline q^a_R \gamma^\mu q^c_R \right)
\left( \overline q^b_R \gamma_\mu (Q^2)^{bc} q^a_R \right) \right.
    \nonumber \\
 &-& \left. 2 \, \left( \overline q^a_L q^c_R \right)
\left( \overline q^b_R (Q^2)^{bc} q^a_L \right)
- 2 \, \left( \overline q^a_R q^c_L \right)
\left( \overline q^b_L (Q^2)^{bc} q^a_R \right) \right] \, .
\ea
Latin indices are for the quark flavours and colour indices are summed
inside the parenthesis.
The meaning of $\alpha_S^{\rm eff}(\mu^2)$ is given at the end of this
section. Penguin diagrams
with photon and gluon interchanged does also
exist but there the photon is at low energies and is thus included
in the low energy part calculated in Section \tref{ld}.

In addition there are also diagrams obtained from inserting
external $v_\mu$, $a_\mu$, $s$ and $p$  currents
on the internal quark lines of the
box- and penguin-like diagrams in Fig. \tref{fig2}.
These do not contribute to the effective action
at order $e^2 p^2$ we consider here.

The factorizable contribution (i.e. the leading order in $1/N_c$
contribution) to the realization of these effective Lagrangian, \rref{box}
and \rref{penguin},
can be easily obtained since,  in that limit,
this Lagrangian is just products of QCD quark currents. The low-energy
realization of QCD quark currents in terms of pseudo-Goldstone bosons and
external sources can be worked out in terms of the
couplings appearing in the QCD chiral Lagrangian. To order $p^2$
this chiral Lagrangian is in \rref{lowest}, at order $p^4$,
it can be found in \rcite{GL85}.

Therefore, to order $p^4$ in the chiral expansion and
leading order in $1/N_c$, the contributions to the
couplings in \rref{e2lowest} and \rref{e2next}, from
integrating out virtual photons with Euclidean momentum larger
than $\mu$ in \rref{empoint} are
\ba
\rlabel{sdcouplings}
C&=& \frac{3}{8} \frac{\alpha_S^{\rm eff}(\mu^2)}{\pi}
 \frac{ F_0^4 B_{0 \rm eff}^2(\mu^2)}{\mu^2} \, ;
\nonumber \\
\tilde K_3& =& \frac{3}{32} \frac{\alpha_S^{\rm eff}(\mu^2)}{\pi}
\frac{F_0^2}{\mu^2} \, ; \hspace*{1cm}
\tilde K_4 = 0 \quad ;  \nonumber \\
\tilde K_5 = \frac{2}{9} \left(\tilde K_3 - \tilde K_6 \right)
\quad &;&\quad
\tilde K_6 = \frac{3}{2} \, \frac{\alpha_S^{\rm eff}(\mu^2)}
{\pi} \frac{ L_5 B_{0 \rm eff}^2(\mu^2)}{\mu^2} \quad ;  \nonumber \\
K_9 &=& - \frac{1}{6} \frac{\alpha_S^{\rm eff}(\mu^2)}{\pi} \,
\frac{ (2 L_8 + H_2) B_{0 \rm eff}^2(\mu^2)}{\mu^2} \, ;
\nonumber \\
K_{10} = K^{\rm QED}_{10} - \frac{9}{2} K_9 \, &;&\quad
K_{11} = K^{\rm QED}_{11} + \frac{3}{4}
\frac{\alpha_S^{\rm eff}(\mu^2)}{\pi} \,
 \frac{ (2 L_8 - H_2) B_{0 \rm eff}^2(\mu^2)}{\mu^2} \, .
\nonumber \\
\ea
All others are zero because of $1/N_c$ counting.
Here, $F_0 \simeq$ $(86\pm 10)$~MeV is the CHPT
chiral limit value  of the pion decay constant $F_\pi$.
The constants $L_i$ and $H_j$ are couplings
of the order $p^4$ strong chiral Lagrangian in \rcite{GL85}.
They reabsorb the UV divergences that appear in CHPT calculations
to order $p^4$.
In particular, the values of the renormalized finite parts 
at a CHPT scale $\tilde \nu=M_\rho$ of the constants we need
in the $\overline{MS}$ scheme used in \rcite{GL85}
are \rcite{reviews}: $L_4= -(0.3 \pm 0.5) \times 10^{-3}$, $L_5 =
(1.4 \pm 0.5) \times 10^{-3}$, $L_6 = -(0.2 \pm 0.3) \times 10^{-3}$,
$L_8 = (0.9 \pm 0.3) \times 10^{-3}$ and
$ 2 L_8 - H_2 = (2.9 \pm 1.0 ) \times 10^{-3}$ \rcite{BPR95}.
The scale dependence of the $L_i$ couplings cancels out when
the next order in $1/N_c$ compared to the one in our calculation is included.
We have chosen the scale $\tilde \nu$ to be a typical
hadronic mass scale. The scales $\tilde \nu$, $\nu$ and $\mu$
are unrelated. The final error assigned to the short-distance
contribution takes into account this $1/N_c$ uncertainty.

The coupling $\tilde K_4$ does not receive
perturbative contributions in QCD and QED
because it modulates a $L^\mu R_\mu$ two-point function.
It can however receive non-perturbative contributions at leading order
in $1/N_c$.
In the case of the ENJL model, these appear
 proportionally to constituent quark masses.

The expressions in Eq. \rref{sdcouplings}
have both order $p^6$ and $1/\mu^4$ corrections.
The product $\alpha^{\rm eff}_S(\mu^2) \,
B_{0 \rm eff}^2(\mu^2)$ has to be understood as
\be
\alpha^{\rm eff}_S(\mu^2) \, B_{0 \rm eff}^2(\mu^2)
\equiv B_0^2(1 {\rm GeV}^2)
\left(\frac{ \dis N_c \alpha_S(1 {\rm GeV}^2)}
{\dis 3}\right)^{-9/11}
{\dis \int_1^\infty}
{\rm d} x \frac{(N_c \alpha_S(x \mu^2)/3)^{2/11}} {x^2}
\ee
and
\be
\alpha^{\rm eff}_S(\mu^2) \equiv {\dis \int_1^\infty}
{\rm d} x \frac{\alpha_S(x \mu^2)}{x^2} \, .
\ee
We use the one-loop large $N_c$ expression
\be
\alpha_S(\mu^2)= \frac{12 \pi}
{11 N_c \ln(\mu^2/ \Lambda_{\rm QCD}^2)}
\ee
with $\Lambda_{\rm QCD}=$ 300 MeV and
$B_0(1 {\rm GeV}^2)=$1.6 GeV  \rcite{BPR95}. This corresponds to using the
large $N_c$ renormalization group to improve the purely perturbative result.

The short-distance contributions to pseudo-Goldstone boson masses
can be found in Appendix \tref{B}.

We can get an estimate of the contributions suppressed in $1/N_c$ 
by keeping the leading in $1/N_c$
contributions from factorization in \rref{box} 
and \rref{penguin} at the quark-gluon  level 
language  but including
the $L_i$ that vanish for $N_c \to \infty$
for the hadronic realization.
This gives an additional set of nonzero terms:
\ba
\tilde K_1 = - \frac{2}{9} \tilde K_2& =& - \frac{2}{3}
\frac{\alpha^{\rm eff}_S(\mu^2)}{\pi}
\frac{L_4 B_{0 \rm eff}^2(\mu^2)}{\mu^2} \, ;
\nonumber \\
 K_7 = - \frac{2}{9} K_8& =& - \frac{4}{3}
\frac{\alpha_S^{\rm eff}(\mu^2)}{\pi}
\frac{L_6 B_{0 \rm eff}^2(\mu^2)}{\mu^2} \, ;
\ea

\subsection{Long-Distance Contribution and Matching}
\rlabel{ld}
We need here the two-point function \rref{empoint} for
external energies up to the Euclidean scale $\mu$. This scale
is expected to be around 1 GeV, thus beyond the applicability of CHPT.
We therefore need to resort to models. We have chosen
the ENJL model for the reasons given below in Sect.
\tref{enjl}.
The low-energy contribution to the two-point function
in \rref{empoint} was calculated already within the same ENJL
model we use here in \rcite{BFP96} for neutral pions.
Here we needed to extend it to any flavour structure.
Since the technique has already been explained several
times\rcite{BP95,BPP96} we will not repeat it here.

Afterwards, we integrate over
the photon momentum in the Euclidean space up to $r_E=\mu$.
This gives us the lower part of the integral in
\rref{cut}.

\subsubsection{The ENJL Model}
\rlabel{enjl}

For recent comprehensive reviews on the NJL \rcite{NJL61}
and the ENJL models \rcite{ENJL75}, see \rcite{japareport,physrep}.
Here, we will only summarize the main features,
notation and reasons why we have chosen this model.
More details and some motivations on the version of the ENJL model
we are using can be found in \rcite{BRZ94,BBR93,BP94}.

The kinetic part of the Lagrangian is given by
\be
\rlabel{QCD}
{\cal L}^{\Lambda_{\rm ENJL}}_{\rm kin} =  
\overline{q} \left\{i\gamma^\mu
\left(\partial_\mu -i v_\mu -i a_\mu \gamma_5 - i G_\mu \right) -
\left({\cal M} + s - i p \gamma_5 \right) \right\} q \, .
\ee
Here summation over colour degrees of freedom
is understood and
we have used the following short-hand notation:
$\overline{q}\equiv\left( \overline{u},\overline{d},
\overline{s}\right)$; $v_\mu$, $a_\mu$, $s$ and
$p$ are external vector, axial-vector, scalar and pseudoscalar
field matrix sources; ${\cal M}$ is the quark-mass matrix;
$G_\mu$ is an external colour source transforming as the gluons
do in QCD. The ENJL model we are using corresponds to
the following Lagrangian
\ba
\rlabel{ENJL1}
{\cal L}_{\rm ENJL} &=& 
{\cal L}_{\rm kin}^{\Lambda_{\rm ENJL}}
+  2 \, g_S \, {\dis \sum_{a,b}} \left(\overline{q}^a_R
q^b_L\right) \left(\overline{q}^b_L q^a_R\right)
\nonumber\\&&
- g_V \, {\dis \sum_{a,b}} \left[
\left(\overline{q}^a_L \gamma^\mu q^b_L\right)
\left(\overline{q}^b_L \gamma_\mu q^a_L\right) +
\left(\overline{q}^a_R \gamma^\mu q^b_R\right)
\left(\overline{q}^b_R \gamma_\mu q^a_R\right)
\right] \,.
\ea
Here $a,b$ are flavour indices, $\Psi_{R(L)} \equiv
(1/2) \left(1 +(-) \gamma_5\right) \Psi$ and
\be
g_V \equiv  {8 \pi^2 G_V(\Lambda_{\rm ENJL})\over N_c 
\Lambda_{\rm ENJL}^2}
\qquad\mbox{,}\qquad g_S
\equiv   {4 \pi^2 G_S(\Lambda_{\rm ENJL})
\over N_c \Lambda_{\rm ENJL}^2}\, .
\ee
The couplings $G_S(\Lambda_{\rm ENJL})$ and $G_V(\Lambda_{\rm ENJL})$ are
dimensionless and ${\cal O}(1)$ in the $1/N_c$ expansion and summation
over colours between brackets in \rref{ENJL1} is understood.

This model has three parameters plus the current light quark masses.
The  first three parameters are $G_S$, $G_V$ and the physical
cut-off $\Lambda_{\rm ENJL}$ of the regularization that we chose to be
proper-time. Although this regulator breaks in general the
Ward identities we impose them by adding the necessary counterterms.
The light quark masses in ${\cal M}$ are fixed in order to
obtain the physical pion and kaon masses in the poles of the
pseudoscalar two-point functions \rcite{BP94}.
The values of the other parameters are fixed from the results of
the fit to low energy effective chiral Lagrangians obtained
in \rcite{BBR93}. They are $G_S \simeq 1.216$, $G_V
\simeq 1.263$, and $\Lambda_{\rm ENJL}
=1.16$ GeV from Fit 1 in that reference.
 Solving the gap equation, we then obtain the
constituent quark masses: $M_u=M_d=275$ MeV and $M_s=427$ MeV.

The model in \rref{ENJL1} is very economical capturing
in a simple fashion
a lot of the observed low and intermediate energy phenomenology. It
has also a few theoretical advantages.

\begin{enumerate}
\item The model in \rref{ENJL1} has the same
symmetry structure as the QCD action
at leading order in $1/N_c$ \rcite{tH74}. In the chiral limit and for $G_S>1$
this model breaks chiral symmetry spontaneously
via the expectation value of the scalar quark-antiquark one-point function
(quark condensate).
\item It has very few free parameters. These are unambiguously
determined from low energy physics involving only pseudo-Goldstone
bosons degrees of freedom\rcite{BBR93}.
\item It only contains constituent quarks. Therefore, all the
contributions to a given process
are uniquely distinguished using only constituent quark diagrams.
Within this model there is thus no possible double counting.
In particular, the constituent quark-loop contribution and what
would be the equivalent of the meson loop contributions in this model,
are of different order in the $1/N_c$ counting.
As described in \rcite{BBR93} this model includes
the constituent-quark loop model as a specific limit.
\item Resummation of fermion-loops automatically produces
a pole in all main spin-isospin channels within the purely constituent
quark picture. This is qualitatively the same as in the
observed hadronic spectrum.
\item It provides a reasonable description of vector and axial
vector meson phenomenology \rcite{PR94}.
\item Some of the short distance behaviour is even the same as in QCD.
For instance, the Weinberg sum rules \rcite{WE67} are satisfied.
These are required in some cases for convergency, for instance
in the Das et al. sum rule \rcite{das}.
\item The major drawback of the ENJL model is the lack of a
 confinement mechanism.
Although one can always introduce an {\em ad-hoc} confining
potential doing the job.
We smear the consequences of this drawback
by working with internal and external
momenta always Euclidean.
\end{enumerate}

\subsubsection{Extraction of the Long-Distance
 Contributions}

The integral over the Euclidean photon momentum is
done at fixed $r_E^2$, i.e. we perform the angular integration
first by a Gaussian procedure. We then fit the
result obtained for the two-point function \rref{empoint}
for a fixed value of the external  momentum $q^2$ to a series of
Chebyshev polynomials in $r_E$.
The remaining integral over $r_E^2$ is then straightforward.
We have choosen the $q^2$ points in the
Euclidean region and near $q^2=0$ where we expect the
artifacts of the model (constituent quark on-shell
effects) to be suppressed.

The lower part of the integral in \rref{cut} has a non-analytical
component too, we have however checked that, in the region of
interest for $|q^2| \le$ 0.2 GeV$^2$ and $r_E\le$ 1 GeV
the Chebyshev polynomials give
a good fit and the non-analytical behaviour is actually very smooth.
We have checked this by calculating the lower part of the integral
within lowest order CHPT with an explicit Euclidean cut-off
$\mu$. For low $\mu$ this agreed with the ENJL calculation.

The two pseudoscalar legs of the
resulting two-point function are then reduced off-shell, for details
of the reduction technique see \rcite{BP94}.
Since we are working in the large $N_c$ limit,
 this has been done for the flavour structures
$\overline  u u$, $\overline d d$, $\overline s s$, $\overline u d$,
$\overline u s$ and $\overline s d$.
 The CHPT expressions for these flavour combinations
in the large $N_c$ limit after the reduction are given in
Appendix \tref{A}.

As an example of the quality we have shown in Fig.
 \tref{fig3},  the reduced two-point function
in the chiral limit as a function of $q^2$ for the integral
in \rref{cut} up to $r_E =$  0.5 GeV for the charged case.
\begin{figure}
\begin{center}
\leavevmode\epsfxsize=12cm\epsfbox{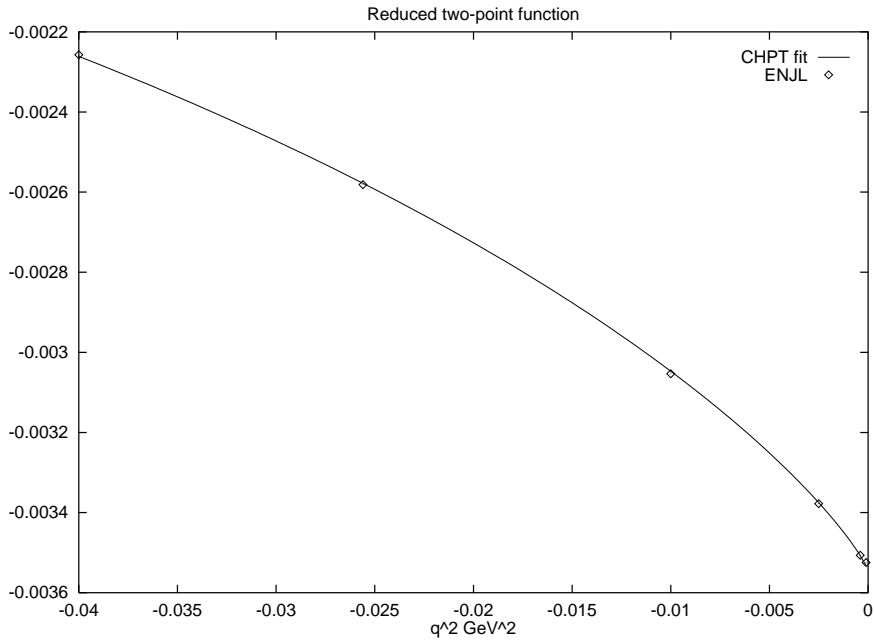}
\end{center}
\caption{\rlabel{fig3}The long distance contribution to the reduced two-point
function as a function of $q^2$ for the integral in
\protect{\rref{cut}} up to $\mu=r_E=$ 0.5 GeV.
The curve is the CHPT fit. All curvature is due to the chiral log.}
\end{figure}
The curvature is purely due to the chiral logarithm. Notice that this is
well reproduced by the ENJL calculation.
Similar good fits were obtained for all the other combinations.
{}From the analytic part of the fit we can then,
 for each flavour case, extract
the combination of coupling constants as given in Appendix \tref{A}.

\subsubsection{Matching}
\rlabel{matching}

Summing the short-distance calculated in the Section
\tref{sd}
and the long-distance obtained in the ENJL model we get the
two-point functions in \rref{twopoi}
with the two pseudoscalar sources reduced.

We have studied the matching of the long- and short-distance contributions
by looking at the stability  in the $\mu$ scale of the
the results. We have plotted the charged pion mass difference
in Figure \tref{figpi}.
\begin{figure}
\begin{center}
\leavevmode\epsfxsize=12cm\epsfbox{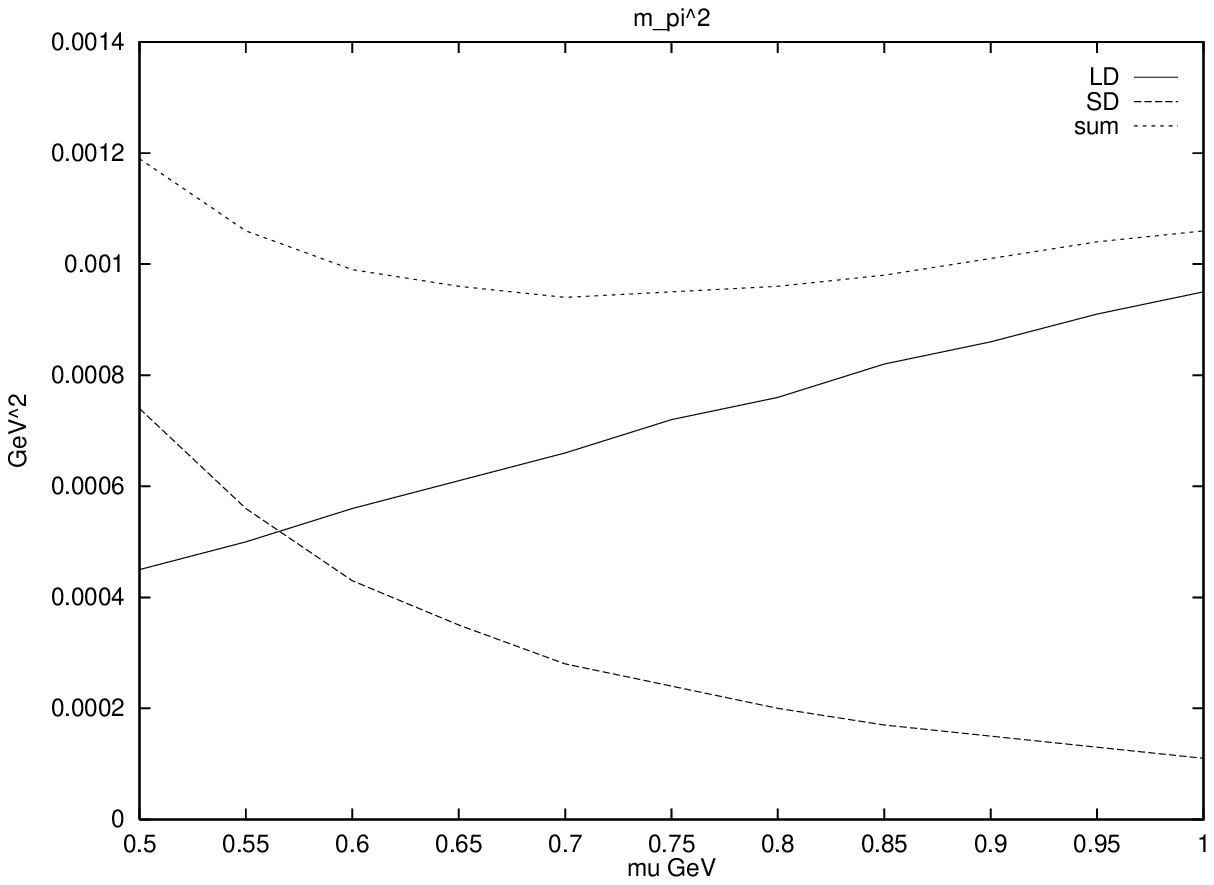}
\end{center}
\caption{\rlabel{figpi}The long-distance (LD), short-distance (SD) and
the sum of the $N_c\to\infty$ results as a function of the matching scale
$\mu$ for $m_{\pi^+}^2|_{\rm EM}$.}
\end{figure}
The matching is quite good  above energies around 
(0.6 $\sim$ 0.8) GeV due to the presence of the photon propagator.
This is because the presence of the photon $1/r^2$ propagator
is enough to cut the high energy contributions.
This happens despite the fact that the
vector and axial-vector propagators obtained within the ENJL
 have only an acceptable behaviour up to around
(0.5 $\sim$ 0.6) GeV in the kaon channel.
We have checked  that this stability (matching) region can be
enlarged just by imposing the correct high-energy behaviour
in the vector two-point functions coupling to the
photon in this case\footnote{For some work where the correct
high-energy behaviour is imposed using QCD-hadron duality
see \rcite{PPR}.}. The predictions for
the couplings remain however mostly unchanged and within
the quoted errors. This is because the presence of
the photon propagator suppresses these contributions.

\section{Results}
\rlabel{results}
\setcounter{equation}{0}

In this section we give the results of our calculation.
In the large $N_c$ limit and $m_u=m_d$ there are four independent
combinations of pseudoscalar two-point functions.
We have calculated the following combinations at
off-shell values of $q^2$ for our analysis:
\begin{enumerate}
\item Neutral case : zero quark masses, equal quark
masses corresponding to the kaon mass
and different quark masses corresponding to the kaon mass.
\item Charged case : zero quark masses, equal quark masses
corresponding to the pion mass,
equal quark masses corresponding to the kaon mass and different
quark masses corresponding to the kaon mass.
\end{enumerate}
To order $e^2 p^2$ there appear five combinations of coupling constants.
 In the long distance case we have worked in the ENJL model to
all orders in the chiral expansion. We will thus also obtain an
estimate of the $e^2 p^4$ and higher corrections
to the long-distance contributions.
{}From this analysis we have got
four of the counterterms of the $e^2 p^2$
Lagrangian in \rcite{urech}. Remember there are ten in the large
$N_c$ limit, where three of them involve external vector
or axial-vector sources. This is presented in Section
\tref{couplings}. We first extract the relevant
corrections to the masses directly.

\subsection{EM corrections to the Masses and Dashen's Theorem}
 \rlabel{dashens}

We take the formulas of Appendix \tref{A} and fit them to
the reduced two-point function of the relevant particle at a fixed value of
$\mu$. We then use the same 
chiral formulas to extrapolate it to the pole.
That way we obtain the long distance contribution to the various masses.
For the short distance we take the results of Section \tref{sd} with
$L_4 = L_6 = 0$ to stay in the $N_c\to\infty$ limit.
The results we obtain are given in Table \tref{tablemass}
where we also quote the
stability region. The contributions for the neutral pion are
are always very small.
\begin{table}
\begin{center}
\begin{tabular}{|c|c|c|c|c|}
\hline
Particle & $m^2_{\rm EM}$ & LD & SD & Stability \\
         & $10^{-3}$  GeV$^2$
     &$10^{-3}$ GeV$^2$&$10^{-3}$ GeV$^2$&$\mu$ in GeV\\
\hline
$\pi^+$ & 0.95 & 0.67 & 0.28 & 0.60 -- 0.85 \\
$K^+$   & 1.93 & 1.47 & 0.46 & 0.65 -- 0.90\\
$K^0$   & $-$0.01 & $-$0.006&$-$0.004& 0.65 -- 0.80\\
$\Delta M^2_{EM}$& 0.98 &0.79 & 0.19 & 0.65 -- 1.00\\
\hline
\end{tabular}
\end{center}
\caption{\rlabel{tablemass}The $N_c\to\infty$ results for the
electromagnetic contributions to the meson masses squared and the violation
of Dashen's theorem. Also quoted are the long- and
short-distance contribution and the stability region.}
\end{table}
The contributions of short and long distance are both
of course $\mu$-dependent.
The numbers given are for the middle of the stability region.
We have also quoted the result directly for the violation of Dashen's
theorem given by
\be
\Delta M_{EM}^2 \equiv \left(m_{K^+}^2-m_{K^0}^2 -m_{\pi^+}^2
+ m_{\pi^0}^2\right)_{\rm EM} \,.
\ee

As an example of the stability we have plotted the
long-distance, the short-distance and the sum of the contributions to
$\left.m_{\pi^+}^2\right|_{\rm EM}$ in Fig. \tref{figpi} and similar for
$\Delta M_{EM}^2$ in Fig. \tref{fig4}. The matching for the other quantities
is not quite as
good but quite acceptable.
\begin{figure}
\begin{center}
\leavevmode\epsfxsize=12cm\epsfbox{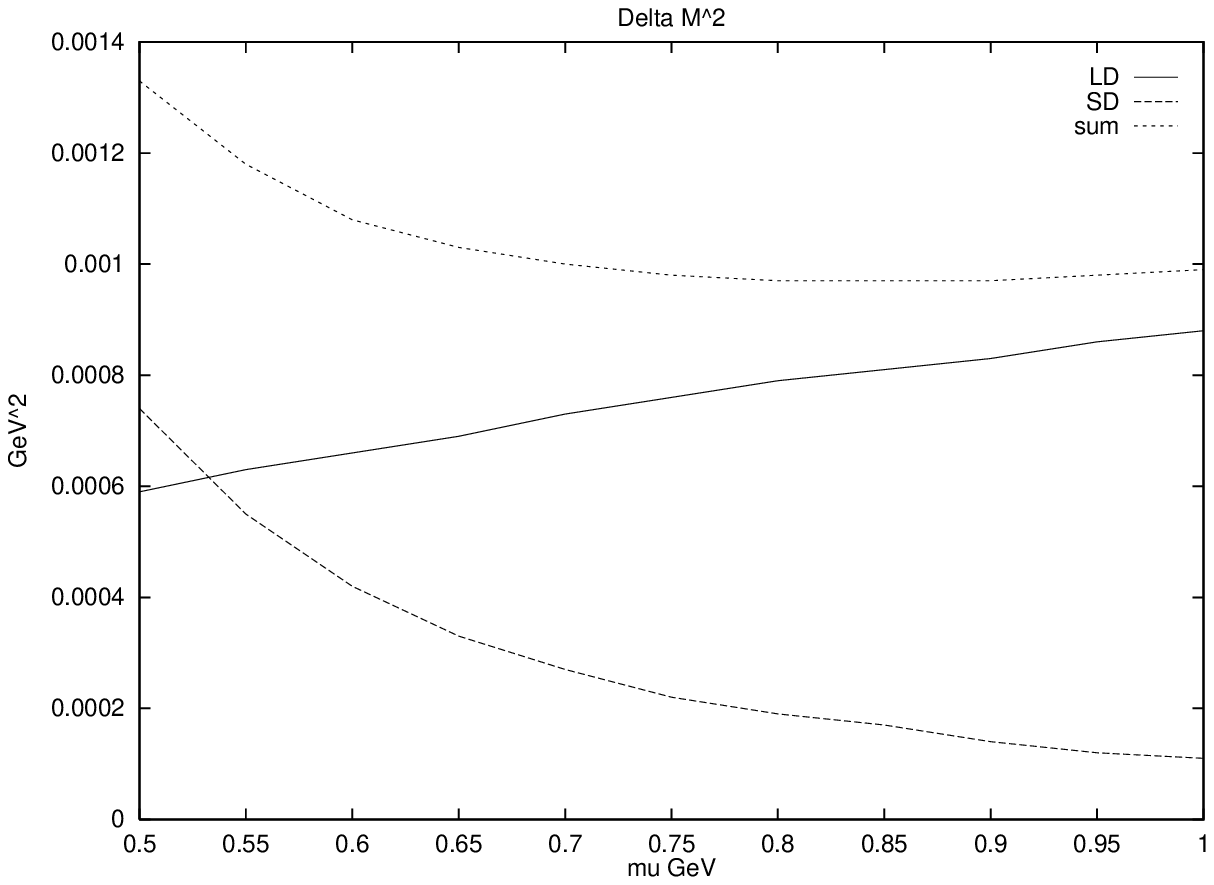}
\end{center}
\caption{\rlabel{fig4}The long-distance (LD), short-distance (SD) and the
sum of the $N_c\to\infty$ results as a function of the matching
scale $\mu$ for $\Delta M_{\rm EM}^2$.}
\end{figure}
There are several reasons for the very good matching
 of $\Delta M_{EM}^2$. First  is that
in this combination, the leading effect the QED quark
mass renormalization
only appears multiplied by pion masses.
In addition, though terms like $m_K^2 \, K_8$
appear for  individual
pseudo-Goldstone boson EM corrections (in the charged pion mass
for instance), they drop out in the combination
$\Delta M_{\rm EM}^2$. Notice that these terms are $1/N_c$
suppressed but appear multiplied by large relative
$m_K^2/m_\pi^2$ factors. The combination $\Delta M_{\rm EM}^2$
has no contributions from counterterms of order $e^2$ $p^2$ $1/N_c$.
This eliminates a potentially large $1/N_c$ uncertainty.
Therefore, EM corrections to pion and kaon masses have larger
uncertainties than the combination $\Delta M_{\rm EM}^2$.

The above numbers are for the current quark masses
defined at the scale $\Lambda =$  1~GeV.
There are contributions from QED running
 of pseudo-Goldstone boson masses both in
the short- and long-distance counterparts. The short-distance ones
are discussed in Section \tref{sd} and can
be obtained from Eqs. \rref{SD1had}. They are the terms proportional
to $K_{10}^{\rm QED}$ and $K_{11}^{\rm QED}$.
As discussed previously, there is a numerical ambiguity
coming from this contribution. This is parametrized
by the UV scale $\Lambda$  in the log dependence
of the short-distance counterpart \rref{SD1}.
The reason for the uncertainty is that the mass definition used here
corresponds to subtracting a QED counterterm corresponding to the
integral from $\Lambda$ till $\infty$. This uncertainty
 we estimate by varying the scale $\Lambda$. Within
the context of the ENJL model (see comments in Section \tref{sd}),
one expects $\Lambda$ to vary between the rho meson mass and
$\Lambda_{\rm ENJL}$, i.e. between 0.8
{\rm GeV} and 1.2 {\rm GeV}, roughly. The
stability region we find for matching between short- and long-distance
contributions is in between 0.6 GeV and 0.9 GeV. We take the 
variation in the contribution 
of $K_{10}^{\rm QED}$ and $K_{11}^{\rm QED}$ by varying the scale 
$\Lambda$ between 0.8 GeV and 1.2 GeV 
as the uncertainty due to the unknown QED counterterm.
 We estimate therefore this uncertainty  in the pseudo-Goldstone
boson masses to be
lower than 1.2 $\times$ $10^{-5}$ GeV$^2$ for  individual pion masses and
cancelling for $(m_{\pi^+}^2-m_{\pi^0}^2)_{\rm EM}$.
 For individual kaon masses it is
smaller than 7.4 $\times$ $10^{-5}$ GeV$^2$. For the combination
$\Delta M_{\rm EM}^2$ it is smaller than
7.5 $\times$ $10^{-6}$ GeV$^2$, thus negligible. In all cases this is smaller
than other uncertainties.

The prediction we get from  our calculation for
$\Delta M_{\rm EM}^2$
due EM virtual corrections and in the large $N_c$ limit is
\be
\rlabel{pionmassdifference}
\Delta M_{\rm EM}^2 = (0.98\pm 0.30) \cdot
10^{-3}~{\rm GeV}^2\,.
\ee

We can now add the leading suppressed $1/N_c$
logarithmic contributions.
These are the log terms
proportional to $C$ in \rcite{urech}.  We include them at a
the CHPT scale $\nu= M_\rho$.  Notice that we are
neglecting the $1/N_c$ contributions from the counterterms.
but they cancel, as mentioned above,  for the combination 
$\Delta M_{\rm EM}^2$.
Using our determination of $C$ from the chiral limit
two-point functions the chiral logs are the second number below
\ba
\rlabel{finresult}
\left.m^2_{\pi^+}\right|_{\rm EM} = &(0.95+0.27)\cdot 10^{-3}~{\rm GeV}^2
&= (1.22\pm0.40)\cdot 10^{-3}~{\rm GeV}^2\nonumber\\
\left.m^2_{\pi^0}\right|_{\rm EM} = &(0.00-0.04)\cdot 10^{-3}~{\rm GeV}^2
&= -(0.04\pm0.01)\cdot 10^{-3}~{\rm GeV}^2\nonumber\\
\left.m^2_{K^+}\right|_{\rm EM} = &(1.93+0.39)\cdot 10^{-3}~{\rm GeV}^2
&= (2.32\pm0.70)\cdot 10^{-3}~{\rm GeV}^2\nonumber\\
\left.m^2_{K^0}\right|_{\rm EM} = &(-0.01+0.00)\cdot 10^{-3}~{\rm GeV}^2
&= -(0.010\pm0.003)\cdot 10^{-3}~{\rm GeV}^2\nonumber\\
\Delta M^2_{EM} = &(0.98+ 0.08)\cdot 10^{-3}~{\rm GeV}^2
&= (1.06\pm0.32)\cdot 10^{-3}~{\rm GeV}^2
\ea
The combination $\Delta M^2_{\em EM}$ is calculated directly,
without using the other results. This is why the error is of
the same order as for the individual contributions.
This is the main result of this work. We confirm a large violation
of Dashen's theorem. Some phenomenological
consequences of it are discussed in Section \tref{quarkmasses}.

In particular the pion mass difference
result should be compared with the experimental mass difference
$m^2_{\pi^+}-m^2_{\pi^0}=$ 1.26 $\cdot  10^{-3}$ GeV$^2$.
As expected the experimental mass difference value
$m_{\pi^+}^2-m_{\pi^0}^2$ is mostly saturated by QED
virtual contributions, with $30\%$ uncertainty though.
The uncertainty here due to not included
$1/N_c$ suppressed counterterm contributions,
is however larger due to the $m_K^2 K_8$ term. An estimate of
its contribution will be discussed in the next section.

\subsection{Determination of Couplings of the $e^2 p^2$ Lagrangian}
\rlabel{couplings}

In this section we give the value of the large $N_c$
couplings in \rref{e2lowest} and \rref{e2next}
that can be determined from our calculation.
Essentially we have fitted the CHPT large $N_c$ results
in Eq. 
\rref{chptres} to the output of our calculation. Being off
mass-shell has allowed to determine one more coupling.
See the comment about the gauge dependence of these couplings
in Section \tref{gauge}. We give the central values at the points
where the stability is best and the errors 
include typical $1/N_c$ error estimates as well as matching
uncertainties. See the comment on matching in Section
\tref{matching}. As customary we take the CHPT scale $\nu$ to be
the rho mass.
The QED short-distance contribution in \rref{SD1had}
is taken at $\Lambda=$ 1 GeV. The matching scale $\mu$
is  always between 0.7 GeV and 0.9 GeV.
For some combinations of couplings we do not get a very good
matching contrary to the masses themselves. The variations 
for $\mu$ between 0.7 GeV and 0.9 GeV are
still within 10\% for most. This seems to be caused by the
large role played by the QED mass renormalization effects.
As an example we have plotted the long distance, the short distance
and the sum for the coefficient $K_{10}$ in Fig. \tref{fig6}
\begin{figure}
\begin{center}
\leavevmode\epsfxsize=12cm\epsfbox{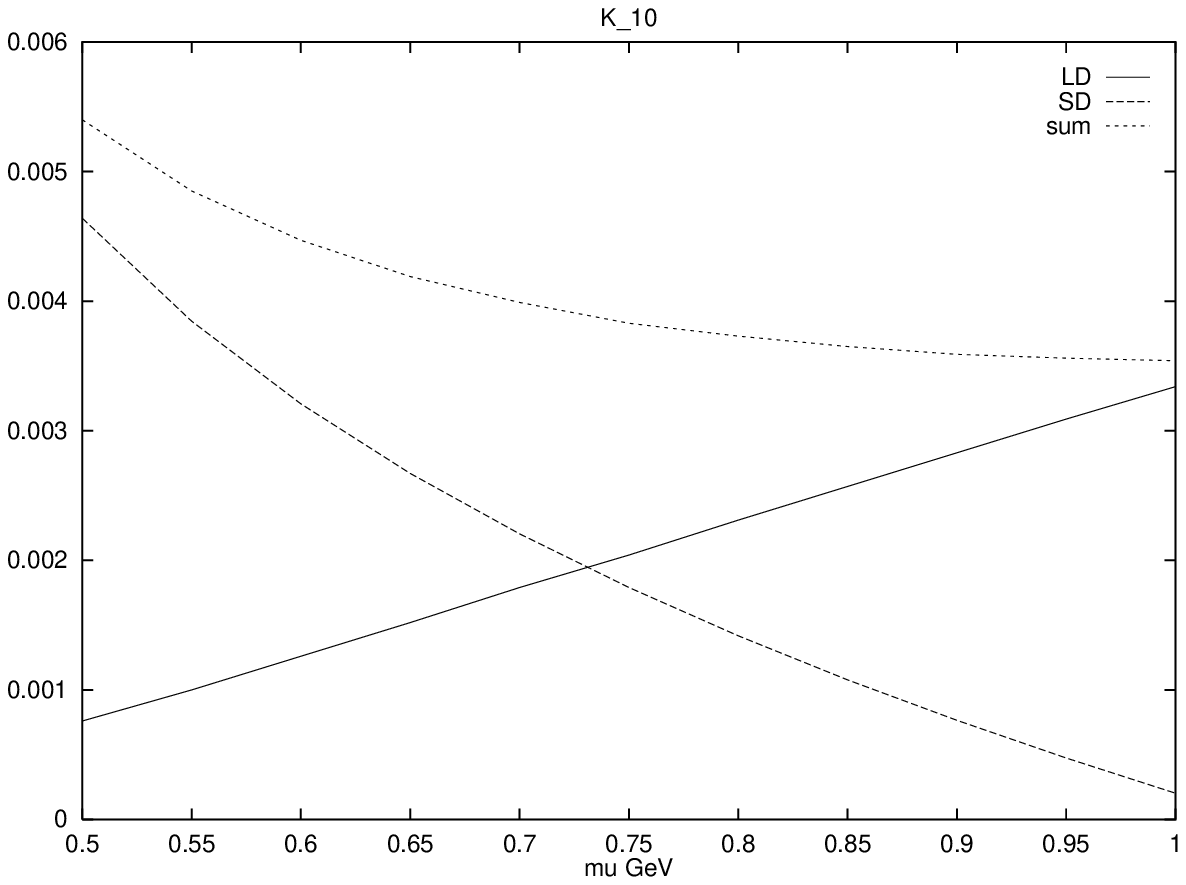}
\end{center}
\caption{\rlabel{fig6}The long-distance (LD), short-distance (SD) and the
sum of the $N_c\to\infty$ results as a function of the matching
scale $\mu$ for $K_{10}(\nu=M_\rho)$.}
\end{figure}
and a combination with very large cancellations in Fig. \tref{fig7}.
\begin{figure}
\begin{center}
\leavevmode\epsfxsize=12cm\epsfbox{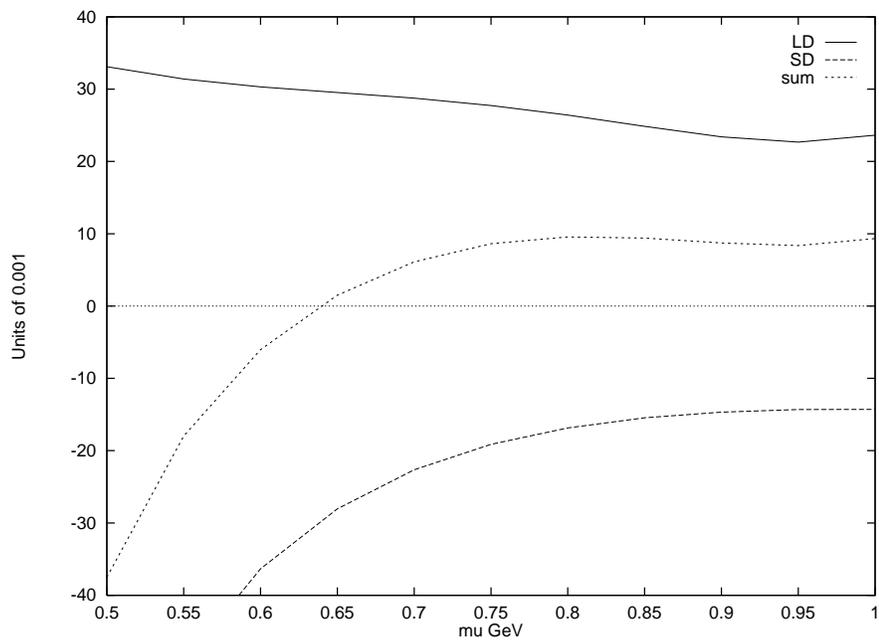}
\end{center}
\caption{\rlabel{fig7}The long-distance (LD), short-distance (SD) and the
sum of the $N_c\to\infty$ results as a function of the matching
scale $\mu$ for $\left(-2 ( 2 \tilde K_3 + \tilde K_4)
+  5 (\tilde K_5 +\tilde K_6) -
10(K_9+K_{10})-18K_{11} \right) (\nu=M_\rho)
$. The units are $10^{-3}$.}
\end{figure}
The other cases are somewhere in between.

{}From  the charged combinations in the chiral limit, i.e. the zero mass
charged pion case, we obtain a good matching and we obtain
\ba
\rlabel{chimass}
&\left.m^2_{\chi^+}\right|_{\rm EM} =
\left[(0.00+0.30+0.67 = 0.97)\pm0.30\right]
\cdot 10^{-3}~{\rm GeV}^2 & \nonumber \\
&{\rm at}\hspace*{0.5cm}\mu=0.85~{\rm GeV}\,,&
\ea
where the first figure
is the short-distance QED contribution, the
second the rest of the short-distance
 and the third is the long-distance part.
The difference with the charged 
pion case is very small. Using consistently the chiral limit
ENJL value\footnote{The chiral limit value of $F_\pi$
is determined in CHPT to be $F_0=(86\pm 10)$~MeV.
Since our low-energy calculation has been done
 within the ENJL model we use here the
value $F_0$ obtained in an ENJL fit to several
low energy observables\protect{\rcite{BBR93}}. Notice
that this value is also compatible with the CHPT value.}
$F_0=$ 89 MeV, we obtain for the $e^2 p^0$
CHPT scale independent coefficient $C$,
\ba
\rlabel{c0}
&C = \left[(0.00+1.3+2.9 = 4.2) \pm 1.5 \right]
\cdot 10^{-5}~{\rm GeV}^4\, &\nonumber \\
&{\rm at}\hspace*{0.5cm}\mu= 0.85 ~{\rm GeV}.&
\ea
The central value  differs by about $30\%$ from earlier determinations,
e.g. \rcite{EGPR89}.
There it was assumed that the full measured pion
mass difference came from the $e^2 p^0$ term
proportional to $C$. Taking into account our error bars
both results are nicely compatible, especially since
the order $e^2 p^2$ contribution  is about 25\%
-- see \rcite{urech,logs} and the results after
Eq. \rref{pionmassdifference}. 

Going off-shell we can get one more combination of
couplings of order $e^2 p^2$, namely
\ba
\rlabel{c1}
& \left(-2 ( 2 \tilde K_3 + \tilde K_4)
+  5 (\tilde K_5 +\tilde K_6) -
10(K_9+K_{10})-18K_{11} \right) (\nu=M_\rho)
=& \nonumber \\
& \left[ (0.85 - 2.53 + 2.64 = 0.96) \pm 0.4 \right]
 \cdot 10^{-2} &\nonumber \\
&{\rm~at~}\hspace*{0.5cm}\mu = 0.8~{\rm GeV}&
\ea
The three first figures are as in \rref{chimass}.

{}From the neutral combinations in the chiral limit
we obtain another combination of couplings, namely
\ba
\rlabel{c2}
& \left(2 \tilde K_3 + \tilde K_4 +
 2 (\tilde K_5 + \tilde K_6) - 4 (K_9 + K_{10}) \right)
 (\nu=M_\rho) = &\nonumber \\
&- \left[ (0.68 - 0.30 + 0.13 = 0.51) \pm 0.3 \right]  \cdot 10^{-2}
& \nonumber \\ & {\rm~at~}\hspace*{0.5cm}\mu = 0.7~{\rm GeV}
\,. &
\ea

Then, by including non-zero current quark masses
we can get one more coupling from the neutral combinations,
 namely
\ba
\rlabel{c3}
&\left(K_9 + K_{10} \right)(\nu=M_\rho)
=\left[ \left( 1.7 + 0.4  + 0.6=2.7\right) \pm 1.0 \right]
\cdot 10^{-3} & \nonumber \\
&{\rm~at~}\hspace*{0.5cm}\mu = 0.7~{\rm GeV} &
\ea
and combining the charged combination with the chiral limit and
the neutral
\ba
\rlabel{c4}
&K_{10}(\nu=M_\rho) =
 \left[ \left( 1.7 +0.5 + 1.8=4.0\right) \pm 1.5 \right]
\cdot 10^{-3} &\nonumber \\
&{\rm~at~}\hspace*{0.5cm}\mu = 0.7~{\rm GeV}&
\ea
from the charged kaon mass EM corrections.
We could have used the $K_{10}$ combination
from the pion mass EM corrections but the
errors due to the subtraction of the chiral limit are larger.
We use it as a check.
Combinations of these four couplings in
\rref{c1}-\rref{c4}  can now
be used for other predictions provided the same  gauge
and $\overline{MS}$ scheme  we use are
used too.

{}From the combinations of couplings above, we can obtain
two combinations which are free of QED uncertainties,
namely,  $ 2 \tilde K_3 + \tilde K_4 +
 2 (\tilde K_5 + \tilde K_6)$ and $K_9$.
In general, we see a strong dependence on the logarithmically
divergent short-distance QED contribution. This makes
more relevant the danger pointed out in Section \tref{sd}
of making a naive chiral power counting here. Fortunately, as
we have seen numerically in the previous section,
when combined with the mass factors and electric charges,
this short-distance QED contribution gives very small
final contribution to the EM mass corrections.

The combinations above are determined directly from simple combinations
of our results. We can combine the numbers in \rref{c1} to \rref{c4}
to obtain as well
\ba
\rlabel{c5}
K_9(\nu=M_\rho) &=& -1.3\cdot 10^{-3}\, ,\\
K_{10}(\nu=M_\rho) &=& 4.0\cdot 10^{-3} \, , \\
\left[2\tilde{K_3}+\tilde{K_4}+2(\tilde{K_5}+\tilde{K_6})\right]
(\nu=M_\rho) &=& 5.7\cdot 10^{-3}\, , \\
\left[2\tilde{K_3}+\tilde{K_4}+4K_{11}\right]
(\nu=M_\rho) &=& -5.0\cdot10^{-3} \, .
\ea

In view of the results we get for the couplings above,
neither short-distance or long distance alone dominate any
of the couplings.
There are in fact large cancellations
in some cases. So that not much can be said about the
couplings we don't get from our calculation.

\section{Comparison with Earlier  Work}
\rlabel{comparison}
\setcounter{equation}{0}

Historically, the soft pion limit was used in the first attempts for
estimating the EM contributions to the pions.
So that, the authors in \rcite{das} arrived to the following expression
for the charged pseudo-Goldstone bosons,
\be
\rlabel{dasrep}
m^2_{\rm EM} = \frac{e^2}{2 F_0^2} i {\dis \int}
\frac{{\rm d}^4 r}{(2 \pi)^4} \, \frac{g_{\mu \nu}
  - \xi r_\mu r_\nu/r^2}{r^2 -i \epsilon}
\left[ \Pi^{(3)  \mu\nu}_{VV} (r^2)
- \Pi^{(3) \mu\nu}_{AA} (r^2) \right] \, ,
\ee
The $\Pi^{(3)}_{VV(AA)}(q^2)$ two-point functions are
\ba
&{\Pi^{(3)\mu\nu}_{VV(AA)}}(q^2) = \nonumber &\\
& i^2 {\dis \int} {\rm d}^4 x e^{iqx} \,
\langle 0 | T \left( \overline q(0) \gamma^\mu (\gamma_5)
{\dis \frac{\dis \lambda_3}{\dis \sqrt 2}} q(0) \right)
\left( \overline q(x) \gamma^\nu (\gamma_5)
{\dis\frac{\dis \lambda_3^\dagger}{\dis \sqrt 2}} q(x) \right)
| 0 \rangle  & \nonumber \\
\ea
for the pions and exchanging the Gell-Mann's flavour SU(3) matrices
$\lambda_3^{(\dagger)}/ \sqrt 2 $ by $[ \lambda_6+(-)i
\lambda_7 ] /2$ for the kaons.
Neutral pseudo-Goldstone bosons get zero contribution in this
limit.
The sum rule in \rref{dasrep} has a good high energy
behaviour in QCD
due to the fulfilling of the Weinberg sum rules (WSRs) \rcite{WE67}.
In fact, in the chiral limit, the first and second WSRs
guarantee its convergency \rcite{das}.
To lowest order in CHPT the integrands in \rref{dasrep}
for kaons and pions are the same, this is again Dashen's theorem.

Attempts to go beyond the approximation in \rref{dasrep}
are in \rcite{urech,neru96,DHW,bau,logs,hans93}.
 Let us compare our results with the ones
obtained in these references.

The chiral logs at $\nu=$ 1 GeV
were used as an estimate of the order of magnitude
of the virtual EM corrections to pseudoscalar Goldstone boson
masses in \rcite{logs}. This is of course a scale dependent
statement, and a conclusion at any particular scale is a dangerous one. 
Only after adding the counterterms the result makes sense.
The present work is devoted to estimate them.

In our final result \rref{finresult},
 we get a relatively large violation of Dashen's theorem,
though  not as large as in \rcite{hans93} and the same
as in \rcite{DHW}.
However, as noticed in  \rcite{bau}, the calculation in
\rcite{DHW} does not have the correct chiral symmetry
behaviour. The short-distance contribution is
here also assumed to be negligible.
We find that this could be the case for scales larger than a few GeV.
There is also work \rcite{perez} improving the estimate made
in \rcite{DHW}. The results found there are compatible with ours.

In Ref. \rcite{bau} some VMD-like estimate in the same line as in
\rcite{DHW} is done. The
matching scale in the resonant saturation in \rcite{bau}
is identified wrongly with the CHPT scale, in our notation
here this is setting $\mu=\nu$. This is very dangerous,
since the chiral logs have in both effective theories
completely different dependence. Also no attempt to make any matching of
the resonant contribution  with the short-distance contribution
was done. These two points make it very hard for us to understand
the meaning of their final result.

Splitting the different
contributions to the corrections of Dashen's lowest order result,
we get
\be
\frac{\left(
m_{K^+}^2-m_{K^0}^2-m_{\pi^+}^2+m_{\pi^0}^2 \right)_{\rm EM}}
{\left(m_{\pi^+}^2-m_{\pi^0}^2 \right)_{\rm exp}}
=\frac{0.98 + 0.08}{1.26} = \,0.84\pm0.25
\ee
where the CHPT scale of the chiral logs is taken
at $\nu=M_\rho$.
The numbers are the leading $N_c$ correction and the $1/N_c$
suppressed chiral logarithms.

Let us now compare this with what we would have gotten from a $e^2 p^2$
calculation using the central values of the $K_i$ counterterms
determined in Section \tref{couplings}
at $N_c\to\infty$:
\ba
\rlabel{large}
\Delta M_{\rm EM}^2
&=& \left[\, 0.94(\gamma K) - 0.16(\gamma \pi) + 0.53(K_i) - 0.31(L_5 C)
\,\right] \cdot 10^{-3} \, {\rm GeV}^{2} \nonumber \\[2mm] &=&
1.00 \cdot 10^{-3} \, {\rm GeV}^{2} \,.
\ea
The four contributions above
are the photon loop contribution to the pion and the kaon,
the contribution from the $K_i$ counterterms and the $L_5 C$
 contribution, respectively.
This should be compared with $0.98 \cdot 10^{-3}$ GeV$^2$
from the full calculation. The agreement is very good, remember
that our nominally $e^2 p^2$ couplings include also
corrections to all orders in CHPT.
Notice also the large cancellation between the $K_i$ and the $L_5 C$
contributions, as mentioned before.
The latter shows the danger of including corrections
of the known constants and fully neglect the others.
The statement is of course somewhat scale dependent.
That the chiral logs give a sizeable part of the total
result  at $\nu=M_\rho$ is a non-trivial dynamical statement
which our calculation answers.

In \rcite{urech,neru96} the order $e^2$ $p^2$ chiral logs
together with some order of magnitude estimate of the
$K_i$ counterterms
was used. The large scale dependence of the logs
only allowed to give a very broad range of results,
of course compatible with the large Dashen's violation we get.

The main difference with the calculation in \rcite{hans93} is the inclusion
of the $1/N_c$ suppressed logarithms and the term proportional
to $L_5$ (last term in \rref{large}).

Recently there has been some lattice QCD results on
the EM contributions to hadrons \rcite{DET} using some
quenched unimproved Wilson action. Their
final result is $\Delta M^2_{\rm EM}=0.64 \cdot 10^{-3}$
 GeV$^2$. Unfortunately only an estimate of part of the
systematic errors due to finite size effects is given there
making difficult the assessment of the result. 
This is particularly relevant  after the recent re-analyses
\rcite{lattice} on the lattice QCD light-quark masses obtained
using  improved Wilson actions  to calculate the same
observables as in \rcite{DET}.
Large ${\cal O}(a)$ lattice spacing effects have been reported
in those works.

There have been several other calculations in the NJL model done.
These were all performed at $G_V=0$, keeping only the scalar four-quark
operators in \rref{ENJL1}. One does however expect already at scales
$\mu$ around 0.5 GeV (as we observe)
non-negligible contributions from spin one structures.
In addition they only treated the pion case, therefore we have not
done a full comparison of our results with those in
\rcite{lemmer}.

We can compare the determination of the couplings from
the previous section with the estimates made in \rcite{urech}
and \rcite{neru,neru96}.
We get that the contribution of the counterterms to
$\Delta M_{\rm EM}^2$
is roughly one order of magnitude smaller than 
the estimate made in those references.

Though the coupling $K_8$ is $1/N_c$ suppressed, its contribution
to the charged pion mass \rcite{urech}
is potentially large due to a factor $m_K^2/m_\pi^2$,
as mentioned before. We can estimate the short distance
 estimate to $K_8$ from \rref{sdcouplings} to be around
$K_8^{\rm SD}(\nu=M_\rho) = - (1.0 \pm 1.4 ) \cdot 10^{-3}$.
An estimate of its total value can be done using the
result in Table \tref{tablemass} for the pion mass difference
and assuming that the
deviation from the experimental result is just due to the 
$1/N_c$
counterterms proportional to $m_K^2$. Notice that we can do this
because we have the complete leading in $1/N_c$ contributions.
In that way we get $K_8 (\nu=M_\rho) = -(0.8 \pm 2.0)
\cdot 10^{-3}$, both compatible with the short-distance estimate
and Zweig's rule.
In \rcite{urech} this same coupling was estimated assuming
that this deviation is dominated completely
by the order $e^2 p^2$
$K_i$ counterterms contributions (both leading  and
next-to-leading in $1/N_c$) proportional to $m_K^2$,
so that they get
$K_8 (\nu=M_\rho) = - (4.0 \pm 1.7) \cdot 10^{-3}$.

\section{Ratios of Light-Quark Masses}
\rlabel{quarkmasses}
\setcounter{equation}{0}

A combination of light current quark masses without the
Kaplan-Manohar \rcite{km} ambiguity, can be obtained using
 the following  relation \rcite{GL85,GL82}
\ba
\rlabel{defQ}
Q^2 &\equiv& \left[ \left( \frac{m_{K^+}^2+ m_{K^0}^2}
{m_{\pi^+}^2+m_{\pi^0}^2} \right)
\,  \left(\frac{m_{K^+}^2+m_{K^0}^2-m_{\pi^+}^2-
m_{\pi^0}^2}{2 \left( m_{K^0}^2-m_{K^+}^2\right)}
\right)\right]_{\rm QCD} \nonumber \\
&=& \, \frac{4 m_s^2 - (m_u+m_d)^2}
{4\left(m_d^2-m_u^2\right)} \left(1 \, + \, {\cal O}
(m_q^2/\Lambda^2_\chi)
 \right) \, .
\ea
As emphasized in \rcite{GL82}, the higher order corrections
to this particular ratio are very suppressed, so it is very
constraining.

In previous sections,
we have computed the virtual EM corrections to the
pseudo-Goldstone bosons masses, the results are in
 Eq. \rref{finresult}.
Subtracting them from the experimental  masses we get
what would be  the QCD
values for those masses. So the result we get for
$Q$ is
\be
\rlabel{finalQ} Q= 22.0 \pm 0.6
\ee
We have calculated the EM corrections at leading order
in $1/N_c$ and to all orders in CHPT. The long
distance contributions are estimated in the ENJL model and
the short-distance contributions in the large $N_c$ limit of QCD.
We also add
the $1/N_c$ suppressed chiral logarithms to order $e^2 p^2$.
The uncertainty corresponds to a 30\% uncertainty on our estimate of
$\Delta M_{EM}^2$.
This should be compared with the lowest order result
\be
Q = 24.2
\ee
using Dashen's theorem.

The $\eta \to \pi^+ \pi^- \pi^0$ decay rate
is inversely proportional to the fourth power of $Q$
\rcite{GL85,DHW}.
Recently there has been  some activity in improving
the prediction  of the proportionality factor
\rcite{KW96,AL96}. Comparing their best estimate
with the experimental data, they get
\ba
Q=22.4 \pm 0.9 \rcite{KW96} \nonumber \\
Q=22.7 \pm 0.8 \rcite{AL96}
\ea
in very good agreement with our result.

To obtain ratios of quark masses themselves requires extra information
beyond CHPT. A very recent
discussion on the consequences of values of $Q$ for the ratios
of the light current quark masses can be found in \rcite{masses}.

\section{Polarizabilities of Pseudo-Goldstone Bosons}
\rlabel{polarizabilities}
\setcounter{equation}{0}

In a previous Letter we calculated the cross-section for
$\gamma\gamma\to\pi^0\pi^0$ to all orders in the chiral expansion and
leading order in $1/N_c$ within the ENJL model\rcite{BFP96}.
As part of the work needed for this paper we need
to calculate the pseudoscalar-pseudoscalar-vector-vector
 (PPVV) four-point function within the same model
also for charged pseudoscalars and any flavour channel.
We can then obtain predictions for
$\gamma\gamma\to\pi^+\pi^-, K^+ K^-, \, \bar K^0 K^0$
to all orders in the chiral expansion and leading in the $1/N_c$
expansion.  The cross-section for the $\gamma\gamma\to\pi^+\pi^-$
process is however
dominated by the Born term and higher corrections to it
are small\rcite{buergi,bijnenscornet}.
The possibility of extracting the pion polarizabilities
from the $\gamma \gamma \to P P^\dagger$ cross-section is studied
in \rcite{DH93}. Combining CHPT with dispersive methods
they arrive to the conclusion that the predicted
polarizabilities are in agreement with the experimental
results. They, however, find low sensitivity in the
cross-sections to the pion polarizabilities (especially
to the neutral ones), so that we cannot expect a precise
determination from the cross-section $\gamma \gamma \to
P P^\dagger$.
The processes $\gamma\gamma\to K^+ K^-, \, K^0 \overline{K^0}$ occur
at a too high center of mass energy for the CHPT predictions
to be reliable.

The polarizabilities for the lowest pseudoscalar mesons
do however fall in the regime where we expect CHPT to work.
 The chiral calculation for the neutral\rcite{BGS94} and
charged\rcite{buergi}  pion polarizabilities have been performed to the
two-loop level. In those works the order $p^6$ counterterms needed
 were estimated by using the resonance saturation model.
Though this model has given good results to order $p^4$
in the strong sector \rcite{EGPR89}, not much is known about its
reliability at ${\cal O}(p^6)$ (and higher). It is therefore important
to compare its predictions with other models, which like the ENJL
model we use, also reproduce the success of the resonance saturation
predictions for the order $p^4$ couplings in the strong sector.
Here we will provide the ENJL model estimates for them.

The polarizabilities for both $P^+$ and $P^0$
are defined by expanding the Compton amplitude
in photon momenta near threshold:
\be
\rlabel{ampdef}
T\equiv 2\left[
 \vec{\epsilon_1}\cdot\vec{\epsilon_2}^*(e^2-4\pi \, m \, \overline \alpha
 \, \omega_1\, \omega_2)
-4\pi \, m \, \overline \beta \, (\vec{q_1}\times\vec{\epsilon_1})
  \cdot(\vec{q_2}\times \vec{\epsilon_2}^*)+\cdots\right]\, ;
\ee
The phase convention we use
can be obtained from this amplitude definition.
Here $m$ is the pseudo-Goldstone boson mass, and
$q \equiv(\omega,\vec{q})$, $\vec{\epsilon}$ are the photon 
momentum and polarization vector, respectively.
  In terms of the relativistic
 amplitudes defined in Eq. (2) of \rcite{BFP96} (we use the same
notation as there), the polarizabilities for both neutral
and charged pseudoscalar bosons are given by
\ba
\rlabel{poldefs}
 \overline \alpha - \overline \beta
  &=& \frac{\alpha}{m} \, {\dis \lim_{s \to 0}} \,
\left(\overline A(s,\nu=s) + 8 m^2 \,
\overline B(s,\nu=s)\right) \nonumber \\
 \overline \alpha + \overline \beta
&=& \frac{\alpha}{m} \,
{\dis \lim_{s \to 0}} \, m^2 \, \overline B (s,\nu=s)\,.
\ea
The barred amplitudes in \rref{poldefs} are the corresponding
amplitudes with the Born contributions using pseudoscalar
propagators to all orders in CHPT  subtracted.

We have numerically calculated the full PPVV four-point function,
 reduced the external pseudoscalar legs (see Ref. \rcite{BP95})
and extracted the $A(s,\nu)$ and $B(s,\nu)$ form factors.
This was done in the same way as was
done in \rcite{BFP96} for the neutral case but now we optimized the
extraction for $s\approx 0$ and $t \approx m^2$.
To order $p^4$ chiral symmetry imposes that $B(s,\nu)=0$.
To the same order  $A(s,\nu)$ also vanishes at large $N_c$
for the neutral pseudo-Goldstone bosons \rcite{bijnenscornet}.
(They have however a non-zero ${\cal O}(p^4)$ contribution from
chiral logs \rcite{bijnenscornet}.) For the charged cases
($\pi^+$ and $K^+$) we get at large $N_c$, i.e. only from
the counterterms,
\be
\rlabel{polp4}
A^{(4)}(s,\nu)= 1.13 \hspace*{0.5cm} {\rm GeV}^{-2}
\ee
at leading order in the $1/N_c$ expansion. This compares well
with the recent determination \rcite{pere}
\be
\rlabel{ampp4}
A^{(4)}(s,\nu) = \frac{8}{F_\pi^2} \, \left( L_9 + L_{10}\right) =
  1.5 \pm 0.2  \hspace*{0.5cm} {\rm GeV}^{-2} .
\ee

CHPT predicts the following counterterm structure
for the form factors  $A(s,\nu)$ and $B(s,\nu)$
\ba
A(s,\nu)_{Count.}&=&A^{(4)}(s,\nu) + \frac{a_1 m^2 + a_2 s}
  {\left(4 \pi F_i\right)^4} \, + \, \cdots \nonumber \, , \\
B(s,\nu)_{Count.}
   &=&\frac{b_1}{\left(4 \pi F_i\right)^4} \,+\,\cdots
\ea
where $F_i$ is $F_\pi$ for pions and $F_K$ for kaons.
In the ENJL model $F_0=89$ MeV, $F_\pi=90$ MeV and
 $F_K=96$ MeV at large $N_c$. The second term in
$A(s,\nu)$ and first in $B(s,\nu)$ are the order
$p^6$ contributions. We have fitted the $A(s,\nu)$
and $B(s,\nu)$ ENJL form factors to a polynomial in $s$
up to order $s^2$. These give always reasonable good fits.
Thus, the $\nu^2$ dependence in the energy region where
 polarizabilities are defined is small.
The results of those fits to the form factors at $A(s,\nu)$ and
$B(s,\nu)$ leading order in $1/N_c$ in terms of $a_1$, $a_2$ and $b_1$
are given in Table \tref{tablecoef} (here we consistently
used the ENJL model values for $F_0$, $F_\pi$ and $F_K$ given above).
We have done it for the chiral limit pseudo-Goldstone boson,
the pion and the kaon with
their appropriate masses, both for the charged and for the
neutral case.
The results we get from the fit
for the coefficients $a_1$, $a_2$ and $b_1$
in Table \tref{tablecoef} include higher than order $p^6$
corrections which are not of the type $s^2$, i.e. mainly mass
corrections.
\begin{table}
\begin{center}
\begin{tabular}{|c|c|c|c|}
\hline
 & $a_1$ & $a_2$ & $b_1$ \\
\hline
$\chi^+$ & -- &6.7 & 0.38 \\
$\pi^+$  & $-$8.7  & 5.9 & 0.38 \\
$K^+$    & $-$5.6 & 15.8 & 0.77 \\
$\chi^0$ & -- & 14.0 & 1.66 \\
$\pi^0$  &$-$23.3&14.9 & 1.69 \\
$K^0$    &$-$13.2&16.9& 1.10 \\
\hline
\end{tabular}
\end{center}
\caption{\rlabel{tablecoef}The dimensionless $a_1$, $a_2$ and $b_1$
coefficients for the chiral limit  pseudo-Goldstone  boson $\chi$,
 the pion and kaon, both for the charged and neutral ones.}
\end{table}
The coefficients in Table \tref{tablecoef} have
a typical error estimate for the $1/N_c$ expansion
of (20 $\sim$ 30) $\%$. In fact the difference
between the coefficients $a_2$ and $b_1$ for the $\pi^+$
and the $K^+$ are higher order corrections.
The coefficients $a_1$ of the charged pion and kaon
have larger uncertainty than the rest since
we get them from subtracting the dominant order $p^4$
contribution from the $A(s,\nu)$ form factor.
{}From the coefficients
of the chiral limit pseudo-Goldstone boson in Table
\tref{tablecoef} and the $a_1$ coefficients for the charged
pion and kaon, one can obtain the ENJL predictions for the six
terms of the order $p^6$ chiral Lagrangian \rcite{p6}
contributing to $\gamma \gamma \to P P^\dagger$ at large $N_c$.

 To get the complete prediction for
the polarizabilities, one has to add to the
counterterm contribution we calculate within the the ENJL
model to all orders in the CHPT expansion, the contributions from chiral
 loop diagrams (order $1/N_c$ in the large $N_c$ counting).
These are for the pions known to two loops \rcite{BGS94,buergi}.
Our final result for the pion and kaons polarizabilities are
 in Tables \tref{tableplus} and  \tref{tableminus}. For the
SU(2) $\overline l_i$ counterterms (see \rcite{GL84} for their definition)
entering the pion chiral log expressions we have consistently used the
ENJL predicted ones, i.e.
\be
\overline l_1 = -1.05, \hspace*{0.5cm}
 \overline l_2 = 5.8, \hspace*{0.5cm}
\overline l_3 = 2.5, \hspace*{0.5cm}
\overline l_4 = 4.3, \hspace*{0.5cm}
\overline l_5 = 14.6, \hspace*{0.5cm}
\overline l_6 = 16.9 \,  .
\ee
\begin{table}
\begin{center}
\begin{tabular}{|c|c|c|c|}
\hline
& $p^6$ and Higher Count. & $p^6$ $\chi$-logs & Total \\
\hline
$\pi^0$  &0.62&0.17&0.79 $\pm$ 0.25 \\
$\pi^+$  &0.14&0.31&0.45 $\pm$ 0.15\\
$K^0$    &1.16&--&--\\
$K^+$    &0.81&--&--\\
\hline
\end{tabular}
\end{center}
\caption{\rlabel{tableplus}The combination of polarizabilities
$\overline \alpha + \overline \beta$ in units of $10^{-4}$
 fm$^3$.
The chiral logs for the kaons (Column 3) are not known.}
\end{table}

\begin{table}
\begin{center}
\begin{tabular}{|c|c|c|c|c|c|}
\hline
& $p^4$ Count.& $p^4$ $\chi$-logs &
  $p^6$ and Higher Count.& $p^6$ $\chi$-logs & Total \\
\hline
$\pi^0$  &0&$-$1.08&$-$0.47&$-$0.31&$-$1.9 $\pm$ 0.6\\
$\pi^+$  &4.53&0&$-$0.25&$-$1.13&3.2 $\pm$ 1.1 \\
$K^0$    &0&--&$-$0.58&--&--\\
$K^+$    &1.27&--&0.07&--&--\\
\hline
\end{tabular}
\end{center}
\caption{\rlabel{tableminus}The combination of polarizabilities
$\overline \alpha - \overline  \beta$ in units of $10^{-4}$
 fm$^3$.
 The chiral logs for the kaons (Columns 3 and  5) are not known.}
\end{table}
The final result in Tables \tref{tableplus} and \tref{tableminus}
contain the counterterm contributions to all orders in CHPT
and large $N_c$.
Notice that the experimental result from \rref{ampp4}
is $(\overline \alpha - \overline \beta)_{{\cal O}(p^4)}=$
$(6.0 \pm 0.8) \cdot 10^{-4}$ fm$^3$.
We can compare the counterterm entries in Tables \tref{tableplus}
and \tref{tableminus} with results obtained
using resonance saturation and other calculations in the ENJL
model\rcite{BB95}.
For the comparison between the resonance saturation
estimates of \rcite{BGS94}
 for the neutral pion case see Table 1 and text in \rcite{BFP96}.
In \rcite{buergi} the following results where obtained from resonance
saturation for the coefficients in Table \tref{tablecoef}
in the case of the $\pi^+$.
\be
a_1=-3.7 \pm 1.65 \, ;
\hspace*{0.5cm} a_2=0.75 \pm 0.65\, ;
\hspace*{0.5cm} b_1=0.45 \pm 0.15 \, .
\ee
Only $b_1$ seems to be in agreement, notice though that
the estimates in \rcite{buergi} don't include the
contributions from scalar and tensor resonances.
 The main part of that work is however  the
two-loop calculation of the charged pion polarizabilities.

 We disagree, as discussed in \rcite{BFP96}, with the way
the order $p^6$ coefficients were obtained in \rcite{BB95}.
We have not compared with the calculations in \rcite{bernard,bajc}
because they work in the ENJL model with $G_V=0$, so the
important (even dominant) effects coming from vector
and axial-vector mesons exchanges are not included.
 For comparison with predictions of other models
for the pion polarizabilities see \rcite{PP}.

  For a recent review of the experimental situation
and data on pion polarizabilities
see \rcite{BGS94,PP} and references therein.

\section{Summary and Conclusions}
\rlabel{conclusions}
\setcounter{equation}{0}

The main aim of this work has been the calculation
of the virtual EM corrections to the masses of
the pseudo-Goldstone bosons. This was motivated by
some recent calculations \rcite{DHW,logs,hans93} where large
corrections to Dashen's theorem were obtained.
This was also supported by recent improved calculations
of the decay rate for $\eta \to \pi^+ \pi^- \pi^0$ \rcite{KW96,AL96}
which included estimates of higher order corrections
using dispersive techniques. At the same time there
appeared some works claiming that
small violations of Dashen's theorem were not excluded
\rcite{urech,neru96,bau}.
We have calculated in the large $N_c$ limit and using
a technique similar to the one in \rcite{BBG87} but
for Green functions off-shell, the virtual
EM mass corrections to pions and kaons, setting
$m_u=m_d$. Our result is
\ba
\Delta M^2_{\rm EM} &=&
\left(m_{K^+}^2-m_{K^0}^2-m_{\pi^+}^2+m_{\pi^0}^2 \right)_{\rm EM}
\nonumber \\ &=&
 (1.06\pm0.32) \cdot 10^{-3}~{\rm GeV}^2
\ea
where we have included the known $1/N_c$ suppressed
chiral logs at order $e^2 p^2$ (they are only 0.08
of that number). The error includes an estimate of
the $1/N_c$ corrections we are missing among other
uncertainties discussed along the text. Notice
that the $1/N_c$ contribution
coming from the order $e^2 p^2$  counterterms cancel
in the combination $\Delta M^2_{\rm EM}$.

Our general conclusion is that a large
violation of Dashen's theorem is quite well
established. At the CHPT scale $\nu=M_\rho$
this is dominated by the photon loop contribution
(both logs and constant pieces).
In fact, our calculation is to all orders in CHPT 
(in the long-distance part) at large
$N_c$, so it includes that contribution to all orders.
The dominance of the photon loop contribution
at $\nu=M_\rho$ is due to a large accidental cancellation
between counterterms of order $e^2 p^2$ of
both types: proportional to $L_i C$ and $K_i$. The inclusion of just part
of them is very dangerous at any scale.
Variation of the CHPT scale in the logs
should be accompanied with the running of the
counterterms which could become eventually important.

A small remark here.
We find some 80 $\%$ correction
to the EM contribution to kaon masses (mainly from the
next-to-leading order). This is very similar and consistent
with what we found in another kaon
self-energy quantity, the so called $B_K$-parameter
\rcite{BP95}. We observe then that two-point function
kaon self-energies from gauge-bosons exchange
have very large higher order CHPT corrections. Notice that in our
approach we are able to make a calculation to {\em all}
orders in CHPT at large $N_c$.

We have obtained also the ratio of
light quark masses $Q^2$ defined in \rref{defQ},
\be
Q = 22.0\pm0.6\,,
\ee
in good agreement with the one found in
\rcite{KW96,AL96}.
Our result supports the very recent scheme  of
ratios of light-quark masses presented by
Leutwyler in \rcite{masses}.
We have also estimated some couplings of the order $e^2 p^2$
effective Lagrangian described by Urech \rcite{urech}.
These were all of the expected order of magnitude. For
values of the combinations obtained here, see Eqs.
 \rref{c1} to \rref{c5}.

We have discussed the ambiguity of the electromagnetic gauge choice in the
definition of these couplings and pointed out how in our approach this is
circumvented. We also discussed how to include the short distance
renormalization needed due to photon loops.

Given the large cancellation observed here between the $K_i$ and the
$L_i\, C$ counterterm contributions to electromagnetic mass differences,
 one should also be careful with estimates of the electromagnetic
corrections to other quantities when only one of these is taken into account.

As a by-product we have also predicted the order $p^6$ terms
which enter in the description of $\gamma \gamma \to
P P^\dagger$ ($P= \pi^+, \pi^0, K^0, K^+$) decays and the
counterterms of
electric and magnetic polarizabilities to all orders
in CHPT for pions and kaons. These predictions were given in Section
\tref{polarizabilities}.

\section*{Acknowledgements}

We would like to thank Peter Gosdzinsky for useful conversations.
J.P. also thanks NORDITA where part of his work was done for hospitality.
The work of J.P. has been supported in part by CICYT (Spain)
under Grant No. AEN96-1718.

\appendix
\section{EM Corrections to Pseudoscalar Two-Point
Functions to ${\cal O}(e^2 p^2)$}
\rlabel{A}
\def\theequation{\Alph{section}.\arabic{equation}}
\setcounter{equation}{0}

In this appendix we give the large $N_c$
 expressions for the electromagnetic
contribution to the pseudoscalar two-point functions in \rref{twopoi}
after reducing
 to order $e^2$ $p^2$ in CHPT. These are finite quantities,
the coupling constants $K_i$ and $\tilde K_i$ are the renormalized
finite parts in the $\overline{MS}$ scheme of \rcite{GL85}
at some scale $\nu$ (this scale has nothing to do with
the scale $\mu$ introduced in \rref{cut} or $\tilde \nu$
in Section \tref{sd}).

We only give here the large $N_c$
expressions off mass-shell, the complete on-shell expressions
to order $e^2$ $p^2$ can be found in \rcite{urech} (with
the translation of couplings in \rref{trans}). These expressions
are the needed ones to obtain the large $N_c$
predictions for the $\tilde K_i$ and $K_i$
counterterms in \rref{e2next} from our calculation.
 We also will use the $\overline q q$ basis
with $q=u,d,s$ quarks which is
more directly related to our large $N_c$
calculation. We use the Feynman gauge for the photon
propagator. The reduced pseudoscalar
two-point functions give the following EM
corrections to the pseudo-Goldstone boson masses
\ba
\rlabel{chptres}
m^2_{\overline u u}(q^2) &=&
- \frac{16e^2}{9} \left[ q^2 \left((2 \tilde K_3 + \tilde K_4)
  \right. \right. \nonumber \\ &+& \left. \left.
 2 (\tilde K_5 + \tilde K_6) - 4 (K_9 + K_{10}) \right)
 + 2  m_\pi^2  (K_9 + K_{10}) \right] \, ; \nonumber \\
 m^2_{\overline d d}(q^2) &=& \frac{1}{4}\,
m^2_{\overline u u}(q^2) \, ;
\nonumber \\
m^2_{\overline s s}(q^2) &=&
- \frac{4e^2}{9} \left[ q^2 \left( (2 \tilde K_3 +
\tilde K_4) \right. \right. \nonumber \\ &+& \left. \left.
 2 (\tilde K_5 + \tilde K_6) -4 (K_9+K_{10})\right) + 2
(2m_K^2-m_\pi^2) (K_9 + K_{10}) \right] \, ;\nonumber \\
m^2_{\pi^+}(q^2)&=&\frac{\dis 2 e^2 C}{\dis F_0^2}
\, \left[ 1- 8 \frac{\dis m_\pi^2}{\dis F_0^2} L_5 \right]
- \frac{\dis e^2}{\dis 16 \pi^2} \left[ m_\pi^2 \left( 3 \ln
\left( \frac{\dis m_\pi^2}{\dis \nu^2} \right) -4 \right) \right.
\nonumber \\ &+& \left.
2 (q^2-m_\pi^2) \left( \left(1 + \frac{\dis m_\pi^2}{\dis q^2}\right)
\ln \left(\frac{\dis m_\pi^2-q^2}{\dis \nu^2}\right)-1-
\frac{\dis m_\pi^2}{\dis q^2}
\ln \left(\frac{\dis m_\pi^2}{\dis \nu^2}\right) \right)
\right] \nonumber \\
&-&\frac{4e^2}{9} \left[ q^2 \left( -2 ( 2 \tilde K_3 + \tilde K_4)
+ 5 (\tilde K_5 +\tilde K_6) - 10 (K_9+K_{10}) -18 K_{11}
\right) \right. \nonumber \\
 &+& \left. m_\pi^2 \left( 5 K_9 -13 K_{10} \right) \right] \, ;
\nonumber \\
m^2_{K^+}(q^2)&=&\frac{\dis 2 e^2C}{F_0^2}
\, \left[ 1- 8 \frac{\dis m_K^2}{\dis F_0^2} L_5 \right]
- \frac{\dis e^2}{\dis 16 \pi^2} \left[ m_K^2 \left( 3 \ln
\left( \frac{\dis m_K^2}{\dis \nu^2} \right) -4 \right) \right.
\nonumber \\
&+& \left. 2 (q^2-m_K^2) \left( \left(1 +
\frac{\dis m_K^2}{\dis q^2}\right)
\ln \left(\frac{\dis m_K^2-q^2}{\dis \nu^2}\right)-1-
\frac{\dis m_K^2}{\dis q^2}
\ln \left(\frac{\dis m_K^2}{\dis \nu^2}\right) \right)
\right] \nonumber \\
&-&\frac{4e^2}{9} \left[ q^2 \left( -2 ( 2 \tilde K_3 + \tilde K_4)
+  5 (\tilde K_5 +\tilde K_6) - 10(K_9+K_{10})-18K_{11} \right)
\right. \nonumber \\ &-& \left.
 3m_\pi^2 (K_9 + K_{10}) + 2 m_K^2
\left( 4 K_9 - 5 K_{10}  \right) \right] \, ; \nonumber \\
 m^2_{\pi^0}(q^2)&=&\frac{1}{2}\left( m^2_{\overline u u}(q^2) +
m^2_{\overline d d}(q^2) \right)\, ; \nonumber \\
 m^2_{K^0}(q^2)&=&\frac{1}{2}\left( m^2_{\overline d d}(q^2) +
m^2_{\overline s s}(q^2) \right)\, .
\ea

\section{Large $N_c$ Short-Distance Contributions}
\rlabel{B}
\setcounter{equation}{0}

In this appendix  we give the short-distance part
of the EM contributions to the pseudoscalar two-point
functions in \rref{twopoi} 
in the large $N_c$ limit. These are the contributions
of the higher than $\mu$ photon modes to the EM corrections
to these pseudoscalar two-point functions
after reducing. (We only give the
independent ones in the large $N_c$ limit.)
\ba
m^2_{\overline u u}(q^2) &=&
- \frac{\dis \alpha \alpha_S^{\rm eff}(\mu^2)}{\dis \mu^2}
 \frac{\dis 4}{\dis 27}
\left[ q^2 \left( 11 F_0^2 + 112 ( L_5 - (2L_8+H_2))
 B_{0 \rm eff}^2(\mu^2) \right) \right. \nonumber \\ &+&
\left. 56 m_\pi^2 (2 L_8 + H_2) B_{0 \rm eff}^2(\mu^2)
\right] +
 \frac{\dis 128 \pi \alpha }{\dis 9} K_{10}^{\rm QED}
\left(2 q^2 - m_\pi^2\right) \, ; \nonumber \\
m^2_{\overline s s}(q^2)&=&
- \frac{\dis \alpha \alpha_S^{\rm eff}(\mu^2)}{\dis \mu^2}
\frac{\dis 1}{\dis 27}
 \left[ q^2 \left( 11 F_0^2 + 112 (L_5 -(2L_8 +H_2))
B_{0 \rm eff}^2(\mu^2) \right)
\right. \nonumber \\ &+& \left. 56 (2 m_K^2 - m_\pi^2)
 (2 L_8 + H_2) B_{0 \rm eff}^2(\mu^2) \right]
\nonumber \\ &+&
 \frac{\dis 32 \pi \alpha }{\dis 9} K_{10}^{\rm QED}
\left(2 q^2 - 2m_K^2+m_\pi^2 \right) \, ; \nonumber \\
m^2_{\pi^+}(q^2)&=& \frac{\alpha \alpha_S^{\rm eff}(\mu^2)}{\mu^2}
\left\{ 3 F_0^2 B_{0 \rm eff}^2(\mu^2)
\left[ 1 - 8 \frac{\dis m_\pi^2}{\dis F_0^2} L_5 \right]
\right . \nonumber \\
&-& \frac{\dis 1}{27} \left[ q^2
\left( -13 F_0^2 + 280 (L_5 - (2L_8+H_2))
B_{0 \rm eff}^2(\mu^2) \right. \right.
\nonumber \\ &-& \left. \left. \left.
 648 (2 L_8 - H_2) B_{0 \rm eff}^2(\mu^2) \right)
- 508 m_\pi^2 (2 L_8 + H_2) B_{0 \rm eff}^2(\mu^2)
   \right] \right\}
\nonumber \\
&-& \frac{16 \pi \alpha}{9}
 K_{10}^{\rm QED} \left(8 q^2 - 13 m_\pi^2
\right) \, ; \nonumber \\
m^2_{K^+}(q^2)&=& \frac{\alpha \alpha_S^{\rm eff}(\mu^2)}{\mu^2}
 \left\{ 3 F_0^2 B_{0 \rm eff}^2(\mu^2)
\left[ 1 - 8 \frac{\dis m_K^2}{\dis F_0^2} L_5 \right]
\right . \nonumber \\
&-& \frac{\dis 1}{27} \left[ q^2
\left( -13 F_0^2 + 280 (L_5 -(2L_8+H_2))  B_{0 \rm eff}^2(\mu^2)
\right. \right. \nonumber \\ &-& \left.
 648 (2 L_8 - H_2) B_{0 \rm eff}^2(\mu^2) \right)
 \nonumber \\ &-& \left. \left.  4 (21 m_\pi^2 +106 m_K^2)
(2L_8 + H_2) B_{0 \rm eff}^2(\mu^2) \right] \right\}
\nonumber \\ &-&
 \frac{16 \pi \alpha}{9} K_{10}^{\rm QED} \left(8 q^2 -
3 m_\pi^2 -10 m_K^2 \right) \, .
\ea

\listoffigures
\end{document}